\documentclass[aps,prl,reprint,superscriptaddress,nofootinbib]{revtex4-2}

\usepackage{amsmath,amssymb,amsfonts,amsthm,mathtools,bm}
\usepackage{microtype}
\usepackage{xcolor}
\usepackage{hyperref}
\hypersetup{colorlinks=true,citecolor=blue!55!black,linkcolor=blue!55!black,urlcolor=blue!55!black,pdftitle={The Quantum Hamming Bound in Arbitrary Local Dimension}}

\newtheorem{theorem}{Theorem}

\newtheorem{corollary}[theorem]{Corollary}
\newcommand{\Vol}{V}
\newcommand{\Kraw}{P}
\newcommand{\Lloyd}{L}

\newcommand{\doilink}[2]{\href{https://doi.org/#1}{#2}}

\begin{document}

\title{The Quantum Hamming Bound in Arbitrary Local Dimension}

\author{Yu-Xuan Zhang}
\affiliation{School of Physics, Nankai University, Tianjin 300071, People's Republic of China}

\author{Jing-Ling Chen}
\email{chenjl@nankai.edu.cn}
\affiliation{Theoretical Physics Division, Chern Institute of Mathematics, Nankai University, Tianjin 300071, People's Republic of China}

\date{\today}

\begin{abstract}
The quantum Hamming bound arises from a direct dimension estimate for exact quantum error-correcting codes. For nondegenerate codes, distinct correctable errors map the code space into orthogonal subspaces, so dimension counting directly gives the bound; degeneracy allows these subspaces to overlap and invalidates the argument. Although linear-programming, asymptotic, and structural approaches have advanced the problem from several directions, the general local-dimension case has remained open. Here we prove the quantum Hamming bound for every local dimension $q\ge3$. Building on the exact two-center Fourier reduction, we develop new overlap estimates for nonbinary Hamming spaces. For $q\ge4$, Jacobi estimates and monotonicity give a uniform gap; qutrits form the boundary case where the gap vanishes and are handled by a filtered correction. This completes the finite-length nonbinary case; together with the existing qubit result, it establishes the quantum Hamming bound for arbitrary local dimension.
\end{abstract}

\maketitle

\textit{Introduction.---}
Quantum error correction links the protection of quantum information against noise to the separation of error subspaces~\cite{Shor1995,CalderbankShor1996,Steane1996,EkertMacchiavello1996,Laflamme1996,KnillLaflamme1997}.  For a nondegenerate code, distinct correctable errors map the code space into mutually orthogonal subspaces.  These subspaces must all fit inside the physical Hilbert space, so a direct dimension count gives a basic constraint.  The code dimension times the number of correctable local errors cannot exceed the dimension of the full Hilbert space; this is the quantum Hamming bound.  It has a direct geometric meaning: the correctable-error spheres must fit together within the available physical state space.  In special nondegenerate cases, perfect quantum Hamming codes attain the bound~\cite{Gottesman1996}.

In contrast to this intuitive nondegenerate picture, the Knill--Laflamme conditions allow distinct correctable errors in a degenerate code to correspond to overlapping error subspaces, invalidating the direct dimension count~\cite{KnillLaflamme1997}.  Motivated by the search for stronger quantum codes, it has long been asked whether the overlap enabled by degeneracy can become large enough to violate the quantum Hamming bound.  Much previous work has focused on stabilizer codes, additive codes, and related classes with additional algebraic structure~\cite{CalderbankRainsShorSloane1998}, and important progress has been made within these settings, but such results do not cover general quantum codes.  Nonadditive codes can achieve better parameters~\cite{RainsHardinShorSloane1997,SmolinSmithWehner2007}.  It is therefore necessary to address the question for quantum codes under the general exact correction conditions.

This question has since been advanced from several directions.  Quantum weight enumerators, shadow enumerators, and linear programming gave code-independent constraints, allowing the quantum Hamming bound to be studied under general correction conditions~\cite{ShorLaflamme1997,RainsWeight1998,Rains1999,AshikhminLitsyn1999}.  Li and Xing further proved that, for fixed distance and sufficiently large block length, the quantum Hamming bound holds~\cite{LiXing2010}.  Subsequent work extended the known range to broader classes of degenerate and $q$-ary quantum codes and several stabilizer settings~\cite{SarvepalliKlappenecker2010,NemecTansuwannont2023}, and ruled out possible violations over large parameter ranges~\cite{Dallas2022}.  Most recently, linear programming combined with node--edge estimates settled the $q=2$ case~\cite{DegeneracyQHB}.  The remaining problem is therefore concentrated in nonbinary local dimensions, requiring pointwise control for every pair $(n,t)$ and every nontrivial exact quantum subspace code.

Here we prove the quantum Hamming bound for exact quantum subspace codes in every local dimension $q\ge3$.  Together with the existing $q=2$ result, this establishes the bound in arbitrary local dimension.  The nonbinary range itself exhibits a clear division by local dimension.  For $q\ge4$, the control of degenerate overlap retains a uniform gap; qutrits sit exactly at the boundary where this gap disappears and therefore require separate treatment.

The key change of viewpoint is that degeneracy forces us to abandon counting several independent error spheres and instead quantify the overlap exactly through intersections of two correctable Hamming balls.  Building on the Fourier reduction and quantum linear-programming techniques of the binary proof~\cite{DegeneracyQHB}, we establish new Jacobi estimates for Hamming-ball intersections in the associated nonbinary Hamming spaces, and combine them with monotonicity in block length to control the ratio between two-center Hamming-ball intersections and the Lloyd response.  For $q\ge4$, these estimates leave a uniform gap, ensuring that the overlap caused by degeneracy cannot violate the quantum Hamming bound.

The qutrit case requires finer treatment.  Shorter block lengths are handled by the Singleton bound together with direct combinatorial estimates, while longer lengths are controlled by a Jacobi comparison; the remaining narrow interval requires a new construction.  We therefore introduce into the linear program a new quadratic-filtered Lloyd square.  Its low-order moment relations preserve the exact normalization required by the quantum Hamming bound while adjusting the sign pattern across the distance shells.  Combined with further estimates on the increments of two-center Hamming-ball intersections, this filtered construction resolves the qutrit case and completes the proof for all nonbinary local dimensions.

For an exact $((n,K,d))_q$ quantum subspace code with $K>1$, write $Q=q^2$.  The quantum Hamming bound reads
\begin{equation}
 K\Vol_t^{(Q)}(n)\le q^n, \quad t=\left\lfloor\frac{d-1}{2}\right\rfloor,
 \label{eq:qhb}
\end{equation}
where $\Vol_t^{(Q)}(n)=\sum_{j=0}^{t}(Q-1)^j\binom nj$ is the size of a radius-$t$ $Q$-ary Hamming ball.

\begin{theorem}[Nonbinary quantum Hamming bound]
\label{thm:nonbinary}
Every exact quantum subspace code $((n,K,d))_q$ with $K>1$ satisfies Eq.~\eqref{eq:qhb} for every local dimension $q\ge3$.
\end{theorem}

The case $t=0$ is immediate; hence assume $t\ge1$.  For $q\ge4$, monotonicity in block length reduces the problem to the critical length $n=4t+1$, while comparison in alphabet size further reduces it to the smallest high-alphabet case $Q=16$.  In this critical regime, the new Jacobi estimates for Hamming-ball intersections provide a uniform gap, which monotonicity then extends to all $q\ge4$.  The qutrit case splits into three length regimes.  Short lengths are handled by the Singleton bound and direct combinatorial estimates, long lengths by Lloyd positivity and Jacobi comparison, while the remaining narrow intermediate interval uses the quadratic-filtered Lloyd square together with estimates on the increments of two-center Hamming-ball intersections to establish the required linear-programming inequalities.  The three regimes cover all qutrit parameters, completing the proof for every $q\ge3$.

\begin{figure}[t]
\centering
\includegraphics[width=0.98\columnwidth]{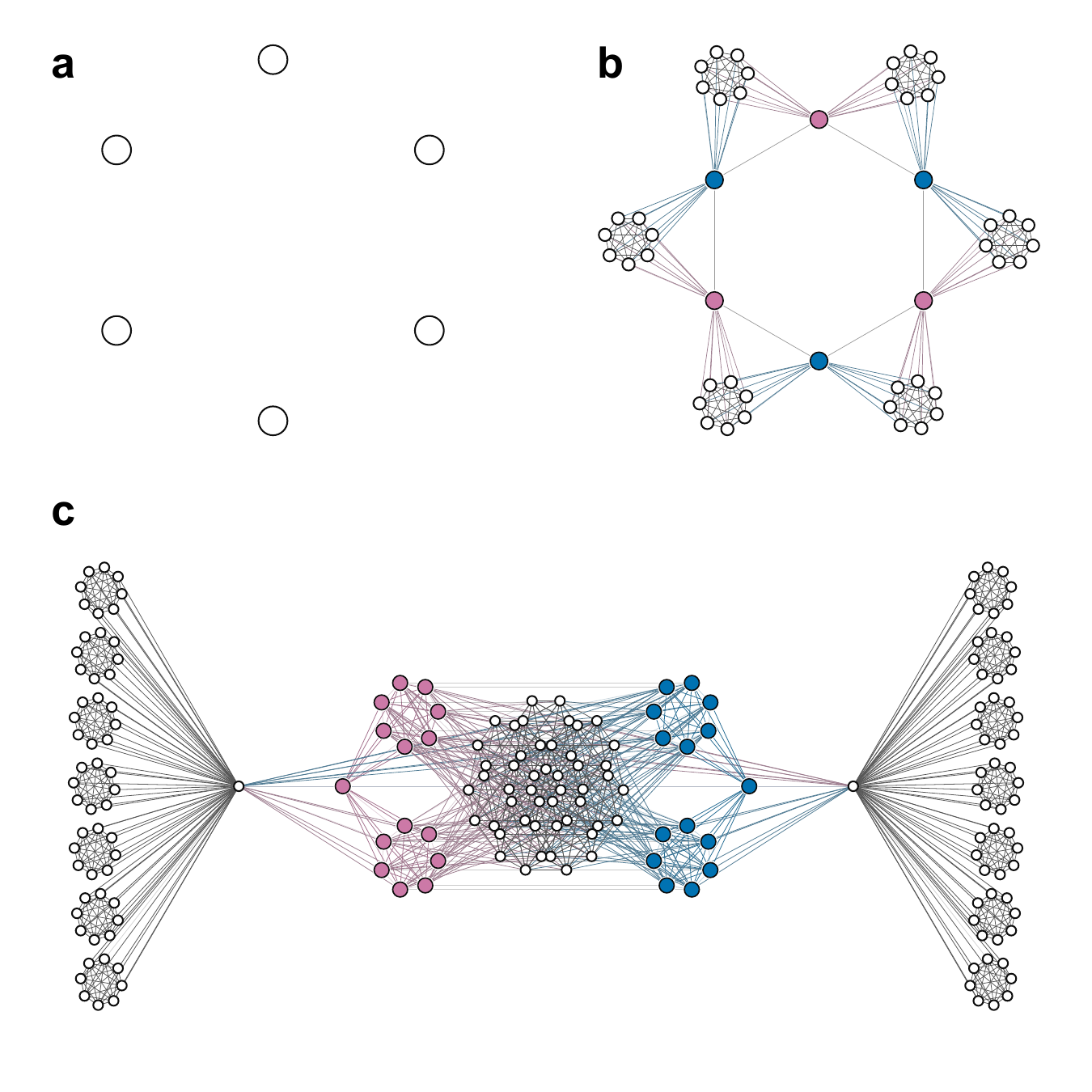}
\caption{Two-center Hamming-ball intersections for $Q=9$, $n=9$, and $t=2$. Panels (a)--(c) show the intersections of two radius-$t$ balls in $H(n,Q)$ at center separations $s=4,3,2$, respectively. Vertices represent $Q$-ary words in the intersection, and edges connect pairs at Hamming distance one. Pink and blue vertices are closer to the first and second center, respectively, while open vertices are equidistant from the two centers.}
\label{fig:two-center-geometry}
\end{figure}

\textit{Two-center Fourier reduction.---}
Let $\Kraw_i^{(Q)}(x;n)$ be the $Q$-ary Krawtchouk polynomial in the standard Hamming association-scheme normalization~\cite{Delsarte1973,MacWilliamsSloane1977}, and define the corresponding Lloyd sum by
\begin{equation}
\begin{aligned}
 \Kraw_i^{(Q)}(s;n) &=\sum_{a=0}^{i}(-1)^a(Q-1)^{i-a}\binom{s}{a}\binom{n-s}{i-a},\\
 \Lloyd_t^{(Q,n)}(x) &=\sum_{i=0}^{t}\Kraw_i^{(Q)}(x;n).
\end{aligned}
\label{eq:kraw-lloyd}
\end{equation}
If $c_t^{(Q)}(n,s)$ denotes the intersection size of two radius-$t$ balls in $H(n,Q)$ whose centers are separated by distance $s$, Fourier inversion in the Hamming association scheme gives
\begin{equation}
 Q^n c_t^{(Q)}(n,s) =\sum_{i=0}^{n}\Lloyd_t^{(Q,n)}(i)^2\Kraw_i^{(Q)}(s;n).
\label{eq:two-center-fourier}
\end{equation}
This identity connects the linear-programming quantities to the geometry of degeneracy through the Fourier transform: the Lloyd square transforms exactly into the intersection of two correctable Hamming balls. The corresponding two-center geometry is illustrated in Fig.~\ref{fig:two-center-geometry} for $Q=9$, $n=9$, and $t=2$. At the largest center separation for which the balls still intersect, $s=2t$, every vertex in the intersection is equidistant from the two centers; as the centers move closer, the intersection grows and develops vertices lying closer to either center.

In the Li--Xing LP theorem~\cite{LiXing2010}, a natural choice is to take the Fourier coefficients as $f_i=\Lloyd_t^{(Q,n)}(i)^2$, for which the ratio at the origin is exactly $Q^n/\Vol_t^{(Q)}(n)$. When $s>2t$, the two radius-$t$ Hamming balls are disjoint, so the corresponding shell sign condition holds automatically with equality. The problem is therefore reduced to the overlap range $1\le s\le2t$, and the Hamming bound follows if
\begin{equation}
 \Lloyd_t^{(Q,n)}(s)>0,  \quad  R_{Q;n,t}(s):= \frac{\Vol_t^{(Q)}(n)c_t^{(Q)}(n,s)}{\Lloyd_t^{(Q,n)}(s)^2}\le1,
 \label{eq:Rdef}
\end{equation}
for every $1\le s\le2t$. Thus the problem in the degenerate case is reduced to a two-center comparison: for each center separation $s$, we only need to control the ratio $R_{Q;n,t}(s)$.

The Supplemental Material (SM) gives the full derivation of the $Q=q^2$ normalization and the Li--Xing reduction above. For the unfiltered Lloyd square, the shell defect to be controlled is $\Lloyd_t^{(Q,n)}(s)^2-\Vol_t^{(Q)}(n)c_t^{(Q)}(n,s)$, which is precisely the quantity treated below in the high-alphabet regime. For $Q=9$ in the bridge window, we retain the same Hamming normalization but introduce a new nonnegative quadratic filter on the Lloyd square. The filter preserves the Hamming ratio at the origin, while the filtered overlapping-shell signs are controlled by balanced increments of the same two-center intersection. Thus the unfiltered and filtered constructions are both built on the same two-center geometry.

\textit{High alphabets.---}
For $q\ge4$, equivalently $Q\ge16$, the two-center comparison arising from the Lloyd square has a uniform margin. For the nontrivial case $K>1$, the quantum Singleton bound gives $n\ge4t+1$. Two monotonicity reductions proved in the SM show that increasing the length beyond the Singleton threshold can only improve the normalized two-center ratio, while increasing the alphabet beyond $Q=16$ can only improve the critical comparison. It therefore suffices to treat the critical line $n=4t+1$ at $Q=16$; on this line, a one-step separation comparison starting from $s=1$ and $s=2$ controls all overlaps.

The key step is to establish a one-step Jacobi comparison in the separation variable. Once the endpoint bounds at $s=1,2$ are known, this comparison controls every larger separation by the value at $s=2$, while the length and alphabet reductions then extend the estimate to all $n\ge4t+1$ and $Q\ge16$. The resulting estimate is stronger than what is required.

\begin{theorem}[High-alphabet half-gap]
\label{thm:halfgap}
For every integer $t\ge1$, every $Q\ge16$, every $n\ge4t+1$, and every $1\le s\le2t$,
\begin{equation}
 \Lloyd_t^{(Q,n)}(s)>0,  \quad R_{Q;n,t}(s)\le\frac12.
 \label{eq:halfgap}
\end{equation}
Consequently, Eq.~\eqref{eq:qhb} holds for every exact quantum subspace code with $K>1$ and $q\ge4$.
\end{theorem}

At the critical line, the interior separation comparison is first reduced to controlling a Hamming-ball intersection ratio. Write $C_s=c_t^{(Q)}(4t+1,s)$ and $I_s=C_s-C_{s+1}$. The node--edge decomposition in the SM organizes $C_s$ and $I_s$ on the same parity chain, giving
\begin{equation}
 \frac{C_s}{I_s}=\frac{v^{\mathsf T}Jv}{v^{\mathsf T}v}\le\lambda_{\max}(J)\le A_s+\frac{3}{\sqrt2}\tau_s,
 \label{eq:high-jacobi}
\end{equation}
where $J$ is the corresponding nonnegative symmetric tridiagonal Jacobi matrix, while $A_s$ and $\tau_s$ control its diagonal and off-diagonal entries, respectively. The two parity Jacobi models and the corresponding entrywise majorants give the final inequality in Eq.~\eqref{eq:high-jacobi}. At the critical length in the high-alphabet regime, this Jacobi bound is combined with a scalar comparison to show that, for $s\ge2$, the normalized two-center ratio is nonincreasing with the center separation. Thus the estimate at $s=2$ propagates along the separation variable to every larger $s$, while $s=1$ is controlled by a separate endpoint estimate. The alphabet comparison and length monotonicity then extend this critical result to all $Q\ge16$ and $n\ge4t+1$. The two endpoint estimates that initiate the separation comparison are
\begin{equation}
\begin{aligned}
 R_{Q;4t+1,t}(1)&\le \frac{4Q(Q-1)}{(3Q-4)^2},\\
 R_{Q;4t+1,t}(2)&\le \frac{16(Q-1)^2(Q^2+6Q-8)}{(3Q-4)^4}.
\end{aligned}
\label{eq:high-endpoints}
\end{equation}
Both are at most $1/2$ for $Q\ge16$. The Jacobi separation comparison gives $R_{Q;4t+1,t}(s+1)\le R_{Q;4t+1,t}(s)$ for $s\ge2$, so the endpoint estimate at $s=2$ propagates to all $2\le s\le2t$ and, together with the endpoint bound at $s=1$, covers the full range $1\le s\le2t$. The length recurrence then propagates the critical estimate to all $n\ge4t+1$. The corresponding algebraic inequalities and scalar rational bounds are proved in the SM.

The high-alphabet proof has two features that will be important below. First, once the critical comparison is established, the estimate is uniform in $t$. Second, the constant $1/2$ is stronger than what the Hamming bound requires: the required threshold is only $1$, so the argument leaves a genuine surplus margin. When $Q$ drops from $16$ to $9$, the same estimates no longer guarantee this surplus, and the qutrit case therefore requires finer control.

\begin{figure*}[t]
\centering
\includegraphics[width=0.98\textwidth]{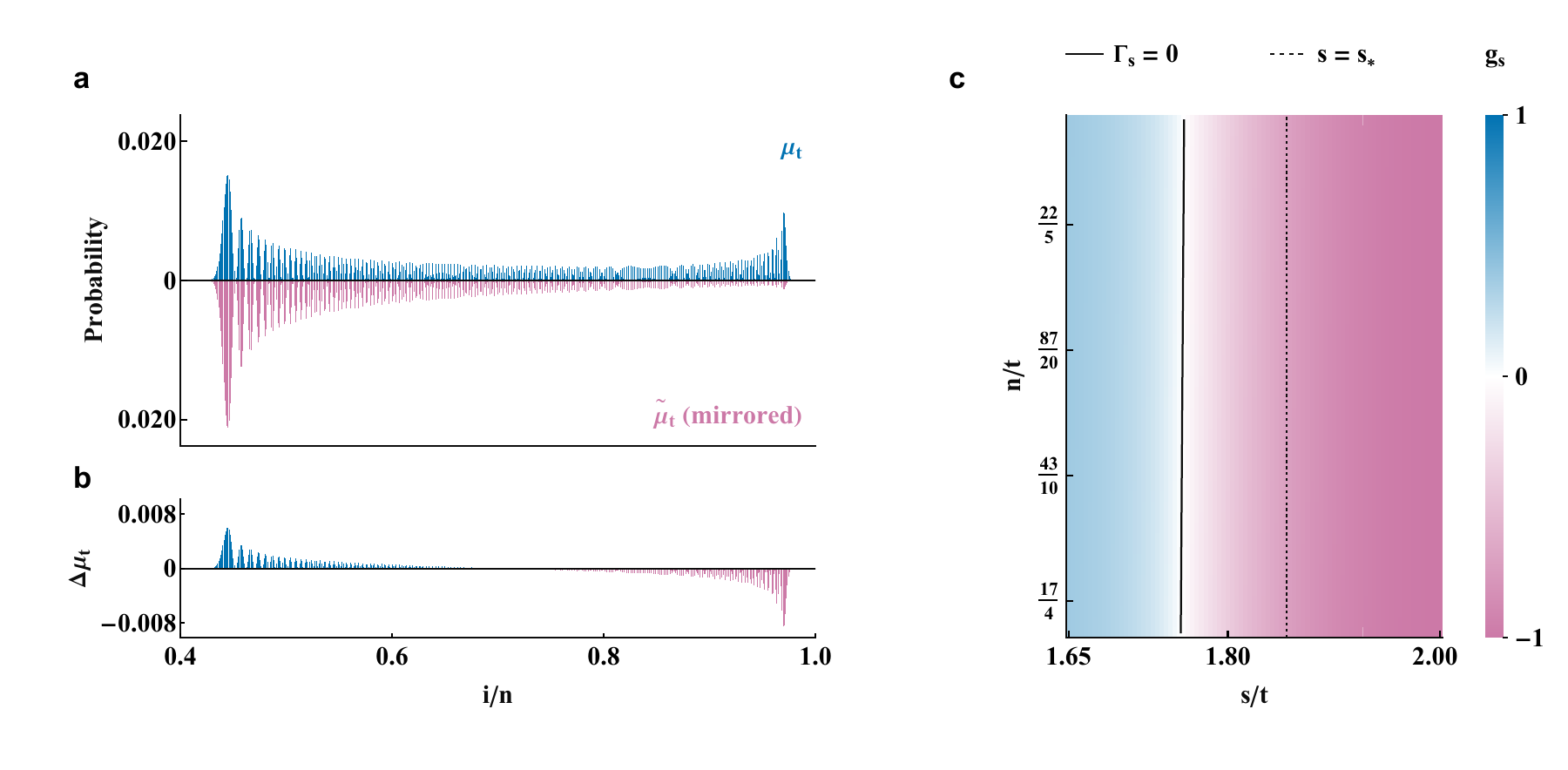}
\caption{Quadratic filtering in the qutrit bridge. (a) The Fourier-shell probability measure induced by the Lloyd square, $\mu_t(i)=8^i\binom{n}{i}\Lloyd_t^{(9,n)}(i)^2/[9^n\Vol_t^{(9)}(n)]$, and its filtered measure $\widetilde{\mu}_t(i)=h_{n,t}(i)\mu_t(i)/h_{n,t}(0)$ for $(n,t)=(1020,240)$; the latter is reflected vertically for visual comparison. (b) The corresponding change in shell weight, $\Delta\mu_t=\widetilde{\mu}_t-\mu_t$. (c) The distance-side response across the bridge for $t=240$, represented by $g_s=\Gamma_s/[h_{n,t}(0)C_s+|\Gamma_s|]$; $g_s$ has the same sign as $\Gamma_s$ and satisfies $|g_s|<1$. The solid curve marks $\Gamma_s=0$, and the dashed line marks the core--tail boundary $s=s_*=371t/200$.}
\label{fig:qutrit-filter}
\end{figure*}

\textit{Qutrit bridge.---}
The only nonbinary case not covered by the high-alphabet half-gap is $q=3$, hence $Q=9$. Here the short- and long-length regimes can still be controlled by the Singleton bound and the original Lloyd square, respectively, while the real difficulty is confined to a narrow intermediate interval. We divide the length line into the short range $17n\le72t$, the long range $84(n-1)\ge373t$, and the bridge between them,
\begin{equation}
\frac{72}{17}t<n<1+\frac{373}{84}t.
\label{eq:bridge-window}
\end{equation}
The short range follows directly from the quantum Singleton bound: $K\le3^{n-4t}$, while $17n\le72t$ gives $\sum_{j=0}^{t}8^j\binom nj\le3^{4t}$, and hence the quantum Hamming bound. In the long range, the original Lloyd-square two-center comparison remains sufficient. The qutrit Lloyd response is already positive for $n-1\ge((27+6\sqrt2)/8)t$; throughout the long range, the preceding Jacobi comparison together with control of adjacent Lloyd values gives $R_{9;n,t}(s+1)\le R_{9;n,t}(s)$ for $2\le s<2t$. It therefore remains only to control $s=1,2$; uniform shell-tail estimates give upper bounds at these two endpoints throughout the long range, covering all admissible $n$. Thus $R_{9;n,t}(s)<1$ for every $1\le s\le2t$. The only remaining lengths are those in the bridge~\eqref{eq:bridge-window}.

This bridge is also the only nonbinary regime in which the linear-programming polynomial must be reconstructed. When $t=1$, the only integer parameter in the bridge is $(t,n)=(1,5)$, for which Eq.~\eqref{eq:Rdef} is verified directly. We therefore assume $t\ge2$ and introduce the quadratic-filtered Lloyd square
\begin{equation}
\begin{aligned}
 f_i&=h_{n,t}(i)\Lloyd_t^{(9,n)}(i)^2,\\
 h_{n,t}(i)&=(n-i)(9i+n+7t-1).
\end{aligned}
\label{eq:filter}
\end{equation}
The multiplier is nonnegative for $0\le i\le n$. More importantly, it is chosen to redistribute the Fourier-shell weight while preserving the exact Hamming normalization. If $F$ denotes the Krawtchouk transform of $f$, the Fourier-shell moment identity proved in the SM gives
\begin{equation}
 \frac{F(0)}{f_0}=\frac{9^n}{\Vol_t^{(9)}(n)}.
 \label{eq:filter-normalization}
\end{equation}
Thus the filtered polynomial enters the Li--Xing linear program with the same origin ratio as the original Lloyd square, leaving only the nonzero distance shells to be controlled.

On the distance side, the structure remains two-centered. Write
\begin{equation}
 C_s=c_t^{(9)}(n,s),\quad I_s=C_s-C_{s+1},\quad r=2t-s.
 \label{eq:bridge-increments}
\end{equation}
The key property of $I_s$ is that this difference preserves the balance between the two centers. The SM proves the exact balanced increment identity
\begin{equation}
 I_s=[u^tv^t](u+v+7uv)^s(1+8uv)^{n-s-1},
 \label{eq:bridge-balanced}
\end{equation}
so $I_s\ge0$; moreover, since $C_{2t+1}=0$,
\begin{equation}
 C_s=\sum_{j=s}^{2t}I_j.
 \label{eq:bridge-suffix}
\end{equation}
At the same time, the Hamming adjacency operator converts the quadratic multiplier in Eq.~\eqref{eq:filter} into a local distance-side expression involving only $C_s$ and the four adjacent increments $I_{s-2},\ldots,I_{s+1}$. Define
\begin{equation}
 \Gamma_s=9^{-n}\sum_{i=0}^{n} f_i\Kraw_i^{(9)}(s;n), \quad H_s=h_{n,t}(s),
 \label{eq:bridge-gamma}
\end{equation}
so that on an overlapping shell the Li--Xing condition becomes
\begin{equation}
 \Phi_s:=H_s\Lloyd_t^{(9,n)}(s)^2-\Vol_t^{(9)}(n)\Gamma_s\ge0.
 \label{eq:bridge-defect}
\end{equation}
On the nonoverlapping shells $s>2t$, it is enough to prove $\Gamma_s\le0$.

Figure~\ref{fig:qutrit-filter} shows the same construction from the Fourier and distance sides. The quadratic multiplier redistributes the Lloyd-square weight across the Fourier shells while preserving the Hamming normalization; after the Krawtchouk transform, the same redistribution changes the sign structure of $\Gamma_s$ on the distance shells. The dashed line marks the boundary between the two estimates used below.

\begin{theorem}[Qutrit bridge estimate]
\label{thm:qutrit-bridge}
For every integer pair $(t,n)$ in the bridge~\eqref{eq:bridge-window} with $t\ge2$, the filtered square~\eqref{eq:filter} satisfies Eq.~\eqref{eq:bridge-defect} for $1\le s\le2t$ and $\Gamma_s\le0$ for $s>2t$.
\end{theorem}

For the overlapping shells, the proof splits according to $r=2t-s$. In the core region
\begin{equation}
 200r\ge29t,
\end{equation}
the original Lloyd square retains enough margin to absorb the correction introduced by the filter. For $t\ge100$, the proof reduces to the two estimates
\begin{equation}
 \Lloyd_t^{(9,n)}(s)^2>2\Vol_t^{(9)}(n)C_s, \quad \Gamma_s\le2H_sC_s.
 \label{eq:bridge-core-pair}
\end{equation}
These immediately imply $\Phi_s>0$. The first follows by propagating $R_{9;n,t}(s)$ through the core, yielding a strict half-gap; the comparison improves with $n$, so the lower endpoint of the bridge controls the entire core. For the second, the adjacent increments in the exact expression for $\Gamma_s$ are transferred onto the same parity chain as $I_s$, reducing the problem to local inequalities. After clearing positive denominators, these inequalities are verified exactly over the full bridge parameter domain. The corresponding transfer identities and exact verification are given in the SM.

Near the edge of the overlap, a more direct mechanism applies. In the tail region $200r<29t$, the balanced increments decay rapidly enough that their ratios control both $C_s$ through Eq.~\eqref{eq:bridge-suffix} and the adjacent terms appearing in $\Gamma_s$. For $t\le j\le2t$, put $R=2t-j$. Then
\begin{equation}
 \frac{I_{j+1}}{I_j}\le \frac{R(2t-R+1)}{7(t-R+1)^2}.
 \label{eq:bridge-tail-ratio}
\end{equation}
For $t\ge225$, this ratio is smaller than $9/160$ throughout the tail. Substitution into the exact increment expression for $\Gamma_s$ gives an upper bound that is increasing in both $n$ and $r$, so it suffices to evaluate it at the upper bridge boundary and the core--tail boundary. This yields $\Gamma_s\le0$ throughout the tail and hence $\Phi_s\ge0$. The remaining finite bridge parameters, namely all overlapping shells with $2\le t<100$ and the tail shells with $100\le t<225$, are verified exactly from the balanced increment formula and the Lloyd recurrence. Finally, for the nonoverlapping shells $s>2t$, the Hamming adjacency operator gives $\Gamma_s\le0$ directly.

This completes the qutrit case. The short range is handled by the Singleton bound and the volume estimate, the long range by the original Lloyd square, and the quadratic-filtered Lloyd square resolves the bridge while preserving the exact Hamming normalization. Combining this with the high-alphabet half-gap proves Theorem~\ref{thm:nonbinary}.

\begin{corollary}[Arbitrary local dimension]
\label{cor:arbitrary}
Combining Theorem~\ref{thm:nonbinary} with the binary finite-length theorem~\cite{DegeneracyQHB}, every exact quantum subspace code $((n,K,d))_q$ with $K>1$ satisfies Eq.~\eqref{eq:qhb} for every local dimension $q\ge2$.
\end{corollary}

\textit{Discussion.---}
Combining the present nonbinary theorem with the existing binary finite-length result~\cite{DegeneracyQHB}, the quantum Hamming bound now holds for exact quantum subspace codes in arbitrary local dimension. For degenerate codes, the error subspaces associated with correctable errors may overlap, so the usual disjoint-sphere counting no longer applies directly. Our result shows that these degeneracy-induced overlaps remain subject to sufficiently strong finite-length constraints, leaving the sphere-packing inequality~\eqref{eq:qhb} unchanged.

The core of the proof is to turn this overlap into a two-center problem in Hamming space. The Fourier transform of the Lloyd square gives the intersection number of two correctable Hamming balls, and the Li--Xing linear program further reduces the quantum Hamming bound to a comparison between the Lloyd response and the two-center intersection. For $q\ge4$, Jacobi estimates for Hamming-ball intersections together with length and alphabet monotonicity yield a uniform half-gap; for $q=3$, the same comparison framework continues to apply but requires a finer treatment of the length regimes. In this way, the difficulty created by degeneracy is concentrated into a definite finite Hamming-geometric quantity, namely the two-center intersection.

This reduction also explains the generality of the result. The two-center intersection number $c_t^{(Q)}(n,s)$ and the Lloyd response $\Lloyd_t^{(Q,n)}(s)$ depend only on the corresponding Hamming association scheme; after the linear-programming reduction, the problem is reduced to estimates among these finite combinatorial quantities. By controlling the maximal overlap that degenerate error subspaces can produce at exact finite length, the proof covers arbitrary quantum codes satisfying the exact quantum error-correction conditions and places degeneracy arising from different code structures within the same finite-length Hamming-geometric framework.

The qutrit bridge further reveals the critical structure of the nonbinary problem. The uniform margin available at high alphabets is no longer sufficient to cover all lengths at $q=3$, but the resulting gap is confined to the narrow bridge between the already controlled short and long regimes. The quadratic-filtered Lloyd square redistributes the Fourier-shell weight while preserving the exact Hamming normalization, and correspondingly adjusts the sign structure on the distance side; the balanced increments then reconnect this new linear-programming condition to the two-center intersection. The bridge is thereby fully controlled, and the quantum Hamming bound continues to hold.

Taken together, the binary, qutrit, and high-alphabet cases reveal a unified finite-length picture: the estimates required to control the overlap vary with the local dimension, while the geometric object underlying them is always the intersection of two correctable Hamming balls. The local dimension determines the analytic strength required to control this intersection, and all regimes ultimately lead to the same sphere-packing bound. The quantum Hamming bound in arbitrary local dimension is therefore reduced to a common two-center Hamming-geometric problem.

\begin{acknowledgments}
This work was supported by the Quantum Science and Technology--National Science and Technology Major Project (Grant No. 2024ZD0301000), and the National Natural Science Foundation of China (Grant No. 12275136).
\end{acknowledgments}

\end{document}


\title{Supplemental Material for ``The Quantum Hamming Bound in Arbitrary Local Dimension''}
\author{Yu-Xuan Zhang}
\affiliation{School of Physics, Nankai University, Tianjin 300071, People's Republic of China}
\author{Jing-Ling Chen}
\affiliation{Theoretical Physics Division, Chern Institute of Mathematics, Nankai University, Tianjin 300071, People's Republic of China}
\date{\today}
\maketitle

\smsection{Fourier reduction and the quantum LP bound}{sec:S1}

Let $Q=q^2$. The case $t=0$ is immediate, since $\Vol_0^{(Q)}(n)=1$ and $K\le q^n$. Hence assume $t\ge1$.

We identify a phase-free $n$-qudit Weyl error with
\begin{equation}
 u=(u_1,\ldots,u_n),\qquad u_k=(a_k,b_k)\in\mathbb Z_q^2,
\end{equation}
and write $\wt u$ for the number of nonzero coordinates. The error labels therefore form the $Q$-ary Hamming space $(\mathbb Z_q^2)^n$. Define
\begin{equation}
 B_t^{(Q,n)}=\{x\in(\mathbb Z_q^2)^n:\wt x\le t\},\qquad \Vol_t^{(Q)}(n)=|B_t^{(Q,n)}|=\sum_{j=0}^{t}(Q-1)^j\binom nj.
 \label{eq:S-volume}
\end{equation}
For $z\in(\mathbb Z_q^2)^n$ with $\wt z=s$, let
\begin{equation}
 c_t^{(Q)}(n,s)=\bigl|B_t^{(Q,n)}\cap(B_t^{(Q,n)}+z)\bigr|.
 \label{eq:S-c-def}
\end{equation}
By Hamming symmetry, this depends only on $s$.

We use the $Q$-ary Krawtchouk normalization
\begin{equation}
 \Kraw_i^{(Q)}(s;n)=\sum_{a=0}^{i}(-1)^a(Q-1)^{i-a}\binom{s}{a}\binom{n-s}{i-a},
 \label{eq:S-kraw}
\end{equation}
with generating function
\begin{equation}
 \sum_{i=0}^{n}\Kraw_i^{(Q)}(s;n)v^i=(1+(Q-1)v)^{n-s}(1-v)^s.
 \label{eq:S-kraw-gen}
\end{equation}
For $u_k=(a_k,b_k)$ and $z_k=(c_k,d_k)$, set $\omega_q=e^{2\pi i/q}$ and
\begin{equation}
 \langle u,z\rangle_s=\sum_{k=1}^{n}(a_kd_k-b_kc_k)\pmod q,\qquad \chi_u(z)=\omega_q^{\langle u,z\rangle_s}.
\end{equation}
The symplectic pairing is nondegenerate, so
\begin{equation}
 \sum_{u\in(\mathbb Z_q^2)^n}\chi_u(z)=\begin{cases}Q^n,&z=0,\\0,&z\ne0.\end{cases}
\end{equation}
For one coordinate, $\sum_{u_k\ne0}\chi_{u_k}(z_k)$ equals $Q-1$ when $z_k=0$ and $-1$ otherwise. Hence, if $\wt z=s$,
\begin{equation}
 \begin{aligned}
  \sum_{i=0}^{n}v^i\sum_{\wt u=i}\chi_u(z)&=\prod_{k=1}^{n}\left(1+v\sum_{u_k\ne0}\chi_{u_k}(z_k)\right)\\
  &=(1+(Q-1)v)^{n-s}(1-v)^s.
 \end{aligned}
\end{equation}
Comparing coefficients with Eq.~\eqref{eq:S-kraw-gen} gives
\begin{equation}
 \sum_{\wt u=i}\chi_u(z)=\Kraw_i^{(Q)}(s;n),\qquad \wt z=s.
 \label{eq:S-shell-sum}
\end{equation}

Define the Lloyd polynomial by
\begin{equation}
 \Lloyd_t^{(Q,n)}(s)=\sum_{j=0}^{t}\Kraw_j^{(Q)}(s;n),
 \label{eq:S-lloyd}
\end{equation}
and let $g=1_{B_t^{(Q,n)}}$. For $h:(\mathbb Z_q^2)^n\to\mathbb C$, use the unnormalized Fourier transform and its inversion formula
\begin{equation}
 \widehat h(u)=\sum_x h(x)\overline{\chi_u(x)},\qquad h(z)=Q^{-n}\sum_u\widehat h(u)\chi_u(z).
\end{equation}
The inversion formula follows from the character orthogonality above. If $\wt u=i$, antisymmetry of the symplectic pairing gives $\overline{\chi_u(x)}=\chi_x(u)$, and Eq.~\eqref{eq:S-shell-sum} gives
\begin{equation}
 \begin{aligned}
  \widehat g(u)&=\sum_{\wt x\le t}\overline{\chi_u(x)}=\sum_{j=0}^{t}\sum_{\wt x=j}\chi_x(u)\\
  &=\sum_{j=0}^{t}\Kraw_j^{(Q)}(i;n)=\Lloyd_t^{(Q,n)}(i).
 \end{aligned}
 \label{eq:S-ball-fourier}
\end{equation}
Define the autocorrelation $A_g(z)=\sum_x g(x)g(x+z)$. If $\wt z=s$, then $A_g(z)=c_t^{(Q)}(n,s)$, while
\begin{equation}
 \begin{aligned}
  \widehat{A_g}(u)&=\sum_{x,z}g(x)g(x+z)\overline{\chi_u(z)}\\
  &=\left(\sum_x g(x)\chi_u(x)\right)\left(\sum_y g(y)\overline{\chi_u(y)}\right)=|\widehat g(u)|^2.
 \end{aligned}
\end{equation}
Fourier inversion therefore gives
\begin{equation}
 \begin{aligned}
  Q^n c_t^{(Q)}(n,s)&=\sum_u|\widehat g(u)|^2\chi_u(z)\\
  &=\sum_{i=0}^{n}\Lloyd_t^{(Q,n)}(i)^2\sum_{\wt u=i}\chi_u(z)\\
  &=\sum_{i=0}^{n}\Lloyd_t^{(Q,n)}(i)^2\Kraw_i^{(Q)}(s;n).
 \end{aligned}
 \label{eq:S-fourier}
\end{equation}
Thus the sequence $\{\Lloyd_t^{(Q,n)}(i)^2\}_{i=0}^n$ has Krawtchouk transform $Q^n c_t^{(Q)}(n,s)$.

The generating function also gives two identities for the Lloyd polynomial. First,
\[
 \Kraw_j^{(Q)}(s;n+1)=\Kraw_j^{(Q)}(s;n)+(Q-1)\Kraw_{j-1}^{(Q)}(s;n),
\]
where $\Kraw_{-1}^{(Q)}=0$. Summing over $0\le j\le t$ gives
\begin{equation}
 \Lloyd_t^{(Q,n+1)}(s)=\Lloyd_t^{(Q,n)}(s)+(Q-1)\Lloyd_{t-1}^{(Q,n)}(s).
 \label{eq:S-lloyd-length}
\end{equation}
For $s\ge1$,
\[
 \sum_{r\ge0}\Lloyd_r^{(Q,n)}(s)v^r=\frac{1}{1-v}\sum_{j\ge0}\Kraw_j^{(Q)}(s;n)v^j=(1+(Q-1)v)^{n-s}(1-v)^{s-1}.
\]
Taking the coefficient of $v^t$ and comparing with the Krawtchouk generating function in length $n-1$ gives
\begin{equation}
 \Lloyd_t^{(Q,n)}(s)=[v^t](1+(Q-1)v)^{n-s}(1-v)^{s-1}=\Kraw_t^{(Q)}(s-1;n-1).
 \label{eq:S-lloyd-shift}
\end{equation}

The Hamming-ball intersection also has a direct count. Fix $y\in B_t^{(Q,n)}\cap(B_t^{(Q,n)}+z)$, where $\wt z=s$. On $\supp z$, let $a$, $b$, and $c$ count the coordinates for which $y_k=0$, $y_k=z_k$, and $y_k\notin\{0,z_k\}$, respectively; outside $\supp z$, let $d$ count the nonzero coordinates of $y$. Then $a+b+c=s$, while
\[
 \wt y=b+c+d,\qquad \wt(y-z)=a+c+d.
\]
Therefore
\begin{equation}
 c_t^{(Q)}(n,s)=\sum_{\substack{a,b,c,d\ge0:\ a+b+c=s\\ b+c+d\le t,\ a+c+d\le t}}\binom{s}{a,b,c}(Q-2)^c\binom{n-s}{d}(Q-1)^d.
 \label{eq:S-c-explicit}
\end{equation}
If the two balls intersect, then $s=\wt z\le\wt y+\wt(y-z)\le2t$. Hence $c_t^{(Q)}(n,s)=0$ for $s>2t$.

We use Li--Xing's quantum linear-programming bound~\cite{LiXing2010} in the following normalization. Let
\begin{equation}
 F(x)=\sum_{i=0}^{n} f_i\Kraw_i^{(Q)}(x;n),\qquad T=\{0,1,\ldots,n\}.
 \label{eq:S-LX-polynomial}
\end{equation}
If $S$ is a nonempty subset of $\{0,\ldots,d-1\}$,
\begin{equation}
 f_i>0\quad(i\in S),\qquad f_i\ge0\quad(i\in T),
 \label{eq:S-LX-coeff-cond}
\end{equation}
and
\begin{equation}
 F(x)\le0\qquad (x\in T\setminus S),
 \label{eq:S-LX-sign-cond}
\end{equation}
then every exact $q$-ary $((n,K,d))$ quantum subspace code satisfies
\begin{equation}
 K\le q^{-n}\max_{x\in S}\frac{F(x)}{f_x}.
 \label{eq:S-LX-bound}
\end{equation}

For $2t\le d-1$, the comparison with the origin can be written in a form that also allows some $f_s$ to vanish. Suppose $f_i\ge0$ for $0\le i\le n$, $f_0>0$, and
\begin{align}
 \frac{F(0)}{f_0}&=\frac{Q^n}{\Vol_t^{(Q)}(n)},
 \label{eq:S-LX-origin}\\
 f_s-\frac{\Vol_t^{(Q)}(n)}{Q^n}F(s)&\ge0\qquad(1\le s\le2t),
 \label{eq:S-LX-shell}
\end{align}
with $F(s)\le0$ for $s>2t$. Set $S=\{0\}\cup\{1\le s\le2t:f_s>0\}$. If $1\le s\le2t$ and $f_s=0$, Eq.~\eqref{eq:S-LX-shell} gives $F(s)\le0$; together with the assumed sign for $s>2t$, this verifies the sign condition on $T\setminus S$. For $s\in S\setminus\{0\}$, Eq.~\eqref{eq:S-LX-shell} gives
\[
 \frac{F(s)}{f_s}\le\frac{Q^n}{\Vol_t^{(Q)}(n)}=\frac{F(0)}{f_0}.
\]
Hence Eq.~\eqref{eq:S-LX-bound} yields
\begin{equation}
 K\le q^{-n}\frac{Q^n}{\Vol_t^{(Q)}(n)}=\frac{q^n}{\Vol_t^{(Q)}(n)}.
 \label{eq:S-qLP-normalization}
\end{equation}

With the coefficient choice
\begin{equation}
 f_i=\Lloyd_t^{(Q,n)}(i)^2,
 \label{eq:S-lloyd-choice}
\end{equation}
Eq.~\eqref{eq:S-fourier} gives
\begin{equation}
 F(s)=Q^n c_t^{(Q)}(n,s).
 \label{eq:S-lloyd-physical-value}
\end{equation}
At the origin, $\Lloyd_t^{(Q,n)}(0)=\Vol_t^{(Q)}(n)$ and $c_t^{(Q)}(n,0)=\Vol_t^{(Q)}(n)$, so
\begin{equation}
 \frac{F(0)}{f_0}=\frac{Q^n}{\Vol_t^{(Q)}(n)}.
 \label{eq:S-lloyd-normalization}
\end{equation}
For $s>2t$, the sign condition holds with equality because $c_t^{(Q)}(n,s)=0$. On $1\le s\le2t$, Eq.~\eqref{eq:S-LX-shell} becomes
\begin{equation}
 \Lloyd_t^{(Q,n)}(s)^2-\Vol_t^{(Q)}(n)c_t^{(Q)}(n,s)\ge0.
 \label{eq:S-lloyd-intersection-coeff}
\end{equation}
For these shells the intersection is nonempty: split $\supp z$ into two sets of size at most $t$, and choose $y$ to equal $z$ on one set and $0$ on the other. Then $\wt y\le t$ and $\wt(y-z)\le t$, so $c_t^{(Q)}(n,s)>0$. Hence \eqref{eq:S-lloyd-intersection-coeff}, if satisfied, also implies $\Lloyd_t^{(Q,n)}(s)\ne0$ and is equivalent to
\begin{equation}
 R_{Q;n,t}(s):=\frac{\Vol_t^{(Q)}(n)c_t^{(Q)}(n,s)}{\Lloyd_t^{(Q,n)}(s)^2}\le1.
 \label{eq:S-LP}
\end{equation}
With $f_i=\Lloyd_t^{(Q,n)}(i)^2$, the Li--Xing reduction is therefore the family of inequalities \eqref{eq:S-LP} on $1\le s\le2t$.

\smsection{Length monotonicity for high alphabets}{sec:S2}

Fix $Q\ge16$ and $m\ge4t+1$. We compare $R_{Q;m,t}(s)$ at the consecutive lengths $m$ and $m+1$.

\begin{lemma}[Length recurrences]
\label{lem:S-length-rec}
For fixed $Q,t,s$, compare lengths $m$ and $m+1$ by taking the two centers to agree in the added coordinate. Then
\begin{align}
 \Vol_t^{(Q)}(m+1)&=\Vol_t^{(Q)}(m)+(Q-1)\Vol_{t-1}^{(Q)}(m),
 \label{eq:S-V-length}\\
 c_t^{(Q)}(m+1,s)&=c_t^{(Q)}(m,s)+(Q-1)c_{t-1}^{(Q)}(m,s),
 \label{eq:S-c-length}\\
 \Lloyd_t^{(Q,m+1)}(s)&=\Lloyd_t^{(Q,m)}(s)+(Q-1)\Lloyd_{t-1}^{(Q,m)}(s).
 \label{eq:S-L-length}
\end{align}
\end{lemma}

\begin{proof}
For the volume, split according to whether the added coordinate is zero or nonzero. The nonzero case has $Q-1$ choices and leaves radius $t-1$ in the first $m$ coordinates. The same split gives the intersection recurrence because the two centers agree in the added coordinate. The Lloyd recurrence is Eq.~\eqref{eq:S-lloyd-length}.
\end{proof}

For $1\le s\le2t$, the two radius-$t$ balls intersect, so $c_t^{(Q)}(m,s)>0$. On a shell for which $\Lloyd_t^{(Q,m)}(s)\ne0$, set
\begin{equation}
 x=\frac{\Vol_{t-1}^{(Q)}(m)}{\Vol_t^{(Q)}(m)},\qquad y=\frac{c_{t-1}^{(Q)}(m,s)}{c_t^{(Q)}(m,s)},\qquad z=\frac{\Lloyd_{t-1}^{(Q,m)}(s)}{\Lloyd_t^{(Q,m)}(s)}.
 \label{eq:S-xyz}
\end{equation}
The one-step comparison follows from
\begin{equation}
 0\le y\le x\le z.
 \label{eq:S-sandwich}
\end{equation}
Indeed, Eq.~\eqref{eq:S-L-length} gives $\Lloyd_t^{(Q,m+1)}(s)=\Lloyd_t^{(Q,m)}(s)(1+(Q-1)z)$, so the sandwich implies that the next Lloyd value is also nonzero. Substituting Eqs.~\eqref{eq:S-V-length}--\eqref{eq:S-L-length} into the definition of $R$ then gives
\begin{equation}
 \frac{R_{Q;m+1,t}(s)}{R_{Q;m,t}(s)}=\frac{(1+(Q-1)x)(1+(Q-1)y)}{(1+(Q-1)z)^2}\le1.
 \label{eq:S-length-factor}
\end{equation}

To prove $y\le x$, decompose $c_t(m,s)$ by the local type \eqref{eq:S-c-explicit}. For fixed $(a,b,c)$ on the support of the displacement, the larger of the two local distances is
\begin{equation}
 h=c+\max(a,b)=s-\min(a,b).
\end{equation}
The remaining $m-s$ coordinates contribute $\Vol_{t-h}^{(Q)}(m-s)$ with a nonnegative coefficient independent of $t$. It is therefore enough to compare the ball ratios term by term. Put $r=t-h$ and $N=m-s$. If $r=0$, the corresponding local type contributes zero to $c_{t-1}^{(Q)}(m,s)$, so there is nothing to prove. Hence assume $r\ge1$.

For any length $M$, define the $j$th shell weight by
\begin{equation}
 w_j^{(Q)}(M)=(Q-1)^j\binom Mj.
\end{equation}
Then
\begin{equation}
 \frac{w_{\ell-1}^{(Q)}(M)}{w_\ell^{(Q)}(M)}=\frac{\ell}{(Q-1)(M-\ell+1)}.
 \label{eq:S-shell-ratio}
\end{equation}
For each $0\le p<r$,
\begin{equation}
 \frac{w_{r-p-1}^{(Q)}(N)}{w_{r-p}^{(Q)}(N)}\le\frac{w_{t-p-1}^{(Q)}(m)}{w_{t-p}^{(Q)}(m)}
 \label{eq:S-adjacent-dom}
\end{equation}
if and only if
\begin{equation}
 h(m-t+p+1)\ge(t-p)(s-h).
\end{equation}
Since $\min(a,b)\le(a+b)/2\le s/2$, we have $h=s-\min(a,b)\ge s-h$. Also, $m\ge4t+1$ implies $m-t+p+1\ge t-p$. Hence \eqref{eq:S-adjacent-dom} holds. Multiplying the adjacent inequalities gives
\begin{equation}
 \frac{w_{r-k}^{(Q)}(N)}{w_r^{(Q)}(N)}\le\frac{w_{t-k}^{(Q)}(m)}{w_t^{(Q)}(m)},\qquad 1\le k\le r.
\end{equation}
Therefore
\begin{equation}
 A_N:=\frac{\Vol_{r-1}^{(Q)}(N)}{w_r^{(Q)}(N)}\le A_m:=\frac{\Vol_{t-1}^{(Q)}(m)}{w_t^{(Q)}(m)}.
\end{equation}
Since the map $A\mapsto A/(1+A)$ is increasing on $[0,\infty)$,
\begin{equation}
 \frac{\Vol_{r-1}^{(Q)}(N)}{\Vol_r^{(Q)}(N)}=\frac{A_N}{1+A_N}\le\frac{A_m}{1+A_m}=\frac{\Vol_{t-1}^{(Q)}(m)}{\Vol_t^{(Q)}(m)}.
\end{equation}
Summing the nonnegative local-type weights, with the $r=0$ terms contributing zero to the numerator, yields
\begin{equation}
 \frac{c_{t-1}^{(Q)}(m,s)}{c_t^{(Q)}(m,s)}\le\frac{\Vol_{t-1}^{(Q)}(m)}{\Vol_t^{(Q)}(m)}.
 \label{eq:S-y-le-x}
\end{equation}
Thus $y\le x$.

The same shell ratio gives a bound for Hamming-ball ratios at any length. For integers $M\ge1$ and $1\le j\le M$, put
\[
 \eta=\frac{j}{(Q-1)(M-j+1)}.
\]
Since the ratio in Eq.~\eqref{eq:S-shell-ratio} is increasing in $\ell$, $w_{\ell-1}^{(Q)}(M)\le\eta w_\ell^{(Q)}(M)$ for $1\le\ell\le j$. Summing gives
\begin{equation}
 \Vol_{j-1}^{(Q)}(M)=\sum_{\ell=0}^{j-1}w_\ell^{(Q)}(M)\le\eta\sum_{\ell=1}^{j}w_\ell^{(Q)}(M)\le\eta\Vol_j^{(Q)}(M),
\end{equation}
and hence
\begin{equation}
 \frac{\Vol_{j-1}^{(Q)}(M)}{\Vol_j^{(Q)}(M)}\le\frac{j}{(Q-1)(M-j+1)}.
 \label{eq:S-shell-tail}
\end{equation}
Taking $M=m$ and $j=t$ gives
\begin{equation}
 x=\frac{\Vol_{t-1}^{(Q)}(m)}{\Vol_t^{(Q)}(m)}\le\frac{t}{(Q-1)(m-t+1)}.
 \label{eq:S-x-upper}
\end{equation}

It remains to bound $z$ from below. By the Lloyd shift identity \eqref{eq:S-lloyd-shift},
\begin{equation}
 z=\frac{\Kraw_{t-1}^{(Q)}(s-1;m-1)}{\Kraw_t^{(Q)}(s-1;m-1)}.
 \label{eq:S-z-kraw}
\end{equation}
Put $N=m-1$ and $u=s-1$, so $N\ge4t$ and $0\le u\le2t-1$. Write $H_i=Q^N(Q-1)^i\binom Ni$, the squared norm of $\Kraw_i^{(Q)}(\cdot;N)$. Then $p_i(u)=(-1)^i\Kraw_i^{(Q)}(u;N)/\sqrt{H_i}$ is an orthonormal Krawtchouk system whose Jacobi matrix has diagonal and off-diagonal entries
\begin{equation}
 b_i=\frac{(Q-1)N-(Q-2)i}{Q},\qquad a_i=\frac{\sqrt{(Q-1)(i+1)(N-i)}}{Q}.
 \label{eq:S-jacobi}
\end{equation}
Therefore the zeros of $\Kraw_t^{(Q)}(u;N)$ are the eigenvalues of the $t\times t$ leading Jacobi truncation. Let $\xi_1$ be the smallest zero. For $t=1$, $\xi_1=b_0$. For $t\ge2$, the $b_i$ decrease with $i$, while $a_i$ increase for $0\le i\le t-2$ because $N\ge4t$. Gershgorin's theorem therefore gives
\begin{equation}
 \xi_1\ge b_{t-1}-2a_{t-2}=\frac{(Q-1)N-(Q-2)(t-1)-2\sqrt{(Q-1)(t-1)(N-t+2)}}{Q}.
 \label{eq:S-zero-lower}
\end{equation}
The same bound holds for $t=1$, with the square-root term equal to zero. Its right-hand side is increasing in $N$ for $N\ge4t$, since its derivative is
\[
 \frac{1}{Q}\left((Q-1)-\sqrt{\frac{(Q-1)(t-1)}{N-t+2}}\right)>0.
\]
It is therefore enough to set $N=4t$. At $N=4t$, the lower bound in \eqref{eq:S-zero-lower} exceeds $2t-1$ precisely when
\[
 (Q-2)t+2(Q-1)>2\sqrt{(Q-1)(t-1)(3t+2)}.
\]
The left-hand side is positive, and after squaring the difference of the squared sides is
\begin{equation}
 (Q^2-16Q+16)t^2+4(Q-1)^2t+4(Q-1)(Q+1)>0,
 \label{eq:S-zero-square}
\end{equation}
which holds for $Q\ge16$ and $t\ge1$. Hence $\xi_1>2t-1$.

It follows that $\Kraw_t^{(Q)}(u;N)$ has no zero on $0\le u\le2t-1$ and is positive there because $\Kraw_t^{(Q)}(0;N)=(Q-1)^t\binom Nt>0$. On this interval the ratio of consecutive Krawtchouk polynomials is increasing. Indeed, since the leading coefficient of $\Kraw_i^{(Q)}(u;N)$ is $(-Q)^i/i!$, the Christoffel--Darboux identity gives
\begin{equation}
 \frac{d}{du}\frac{\Kraw_{t-1}^{(Q)}(u;N)}{\Kraw_t^{(Q)}(u;N)}=\frac{QH_{t-1}}{t\,\Kraw_t^{(Q)}(u;N)^2}\sum_{i=0}^{t-1}\frac{\Kraw_i^{(Q)}(u;N)^2}{H_i}>0.
 \label{eq:S-kraw-ratio-monotone}
\end{equation}
Therefore, using $u=s-1\ge0$,
\begin{equation}
 z\ge\frac{\Kraw_{t-1}^{(Q)}(0;m-1)}{\Kraw_t^{(Q)}(0;m-1)}=\frac{t}{(Q-1)(m-t)}.
 \label{eq:S-z-lower}
\end{equation}
Together with \eqref{eq:S-x-upper}, this gives
\begin{equation}
 x\le\frac{t}{(Q-1)(m-t+1)}<\frac{t}{(Q-1)(m-t)}\le z,
\end{equation}
so the sandwich \eqref{eq:S-sandwich} holds. The same first-zero argument also shows $\Lloyd_t^{(Q,m)}(s)=\Kraw_t^{(Q)}(s-1;m-1)>0$ for $1\le s\le2t$. Hence \eqref{eq:S-length-factor} yields
\begin{equation}
 R_{Q;m+1,t}(s)\le R_{Q;m,t}(s),\qquad Q\ge16,\quad m\ge4t+1,\quad 1\le s\le2t.
 \label{eq:S-length-monotone}
\end{equation}
Iterating from $m=4t+1$ to $m=n-1$ gives
\begin{equation}
 R_{Q;n,t}(s)\le R_{Q;4t+1,t}(s),\qquad n\ge4t+1.
\end{equation}
Thus, for $Q\ge16$, the bound at every $n\ge4t+1$ follows from the critical case $R_{Q;4t+1,t}(s)\le1$ for $1\le s\le2t$.

\smsection{Jacobi bounds for Hamming-ball intersections}{sec:Sjacobi}

Fix $Q\ge3$, $t\ge1$, $n\ge2t+1$, and $0\le s\le2t$. Put $C_s=c_t^{(Q)}(n,s)$ and $I_s=C_s-C_{s+1}$. Let
\begin{equation}
 a(u,v)=u+v+(Q-2)uv,\qquad b(u,v)=1+(Q-1)uv.
 \label{eq:S-jacobi-ab}
\end{equation}
If $[u^{\le t}v^{\le t}]$ denotes the sum of coefficients of $u^av^b$ with $0\le a,b\le t$, then the two-ball intersection has the coefficient form
\begin{equation}
 C_s=[u^{\le t}v^{\le t}]a(u,v)^s b(u,v)^{n-s}.
 \label{eq:S-jacobi-C-coeff}
\end{equation}
Indeed, on each of the $s$ separated coordinates the three possibilities contribute $u$, $v$, and $(Q-2)uv$, while each of the remaining $n-s$ coordinates contributes $1+(Q-1)uv$. Since $b-a=(1-u)(1-v)$, subtraction at consecutive separations gives
\begin{equation}
 I_s=[u^{\le t}v^{\le t}]a^s b^{n-s-1}(1-u)(1-v)=[u^tv^t]a^s b^{n-s-1}.
 \label{eq:S-jacobi-I-coeff}
\end{equation}
The last equality is the two-variable telescoping identity $[u^{\le t}v^{\le t}](1-u)(1-v)P=[u^tv^t]P$ for a polynomial $P$ with nonnegative powers.

Put
\begin{equation}
 r=2t-s,\qquad N=n-s-1.
 \label{eq:S-jacobi-rN}
\end{equation}
Since $n\ge2t+1$, we have $N\ge r$. Write $w_j^{(Q)}(M)=(Q-1)^j\binom Mj$ for the $j$th Hamming shell weight. In the coefficient of $u^tv^t$ in \eqref{eq:S-jacobi-I-coeff}, let $c$ be the number of $(Q-2)uv$ factors chosen from $a^s$. The remaining $m_c=s-c$ separated coordinates must split equally between the $u$ and $v$ terms, so $c\equiv s\pmod2$. Define
\begin{equation}
 \mathcal C_s=\{c:0\le c\le\min(s,r),\ c\equiv s\pmod2\},\qquad m_c=s-c,\qquad \ell_c=\frac{r-c}{2},
 \label{eq:S-jacobi-Cset}
\end{equation}
and, for $c\in\mathcal C_s$,
\begin{equation}
 D_c=\binom{s}{c}(Q-2)^c\binom{m_c}{m_c/2}w_{\ell_c}^{(Q)}(N).
 \label{eq:S-jacobi-Dc}
\end{equation}
All $D_c$ are positive, and expanding \eqref{eq:S-jacobi-I-coeff} gives
\begin{equation}
 I_s=\sum_{c\in\mathcal C_s}D_c.
 \label{eq:S-jacobi-I-nodes}
\end{equation}
The set $\mathcal C_s$ is nonempty: it contains $0$ when $s$ is even, while for odd $s\le2t$ one has $s\le2t-1$ and hence $1\in\mathcal C_s$. Therefore $I_s>0$.

Before grouping the full intersection by parity, we record its direct separated-coordinate expansion. If $c$ of the $s$ separated coordinates contribute $(Q-2)uv$ and, among the remaining $s-c$ coordinates, $p$ contribute $u$, then the two distances contributed on the separated coordinates are $p+c$ and $s-p$. The remaining $n-s=N+1$ coordinates may therefore have weight at most $t-\max(p+c,s-p)$. Hence
\begin{equation}
 c_t^{(Q)}(n,s)=\sum_{\substack{a,b,c,d\ge0:\\ a+b+c=s\\ b+c+d\le t,\ a+c+d\le t}}\binom{s}{a,b,c}(Q-2)^c\binom{n-s}{d}(Q-1)^d.
 \label{eq:S-jacobi-c-explicit}
\end{equation}
We now regroup this exact sum according to the parity of $s-c$ and the imbalance between the numbers of $u$ and $v$ terms. For $c\in\mathcal C_s$, an imbalance $2j$ between the numbers of $u$ and $v$ terms leaves outside radius $\ell_c-j$, hence
\begin{equation}
 \alpha_c=\sum_{j=0}^{\min(m_c/2,\ell_c)}(2-\delta_{j0})\frac{\binom{m_c}{m_c/2-j}}{\binom{m_c}{m_c/2}}\frac{\Vol_{\ell_c-j}^{(Q)}(N+1)}{w_{\ell_c}^{(Q)}(N)}.
 \label{eq:S-jacobi-alpha}
\end{equation}
When the number of $(Q-2)uv$ factors has the opposite parity, write it as $e$ and define the ordinary-edge set
\begin{equation}
 \mathcal E_s=\{e:1\le e\le\min(s,r-1),\ e\not\equiv s\pmod2\},\qquad \rho_e=\frac{r-e-1}{2}.
 \label{eq:S-jacobi-Eset}
\end{equation}
For $e\in\mathcal E_s$, put
\begin{align}
 B_e&=\binom{s}{e}(Q-2)^e\binom{s-e}{(s-e-1)/2}w_{\rho_e}^{(Q)}(N),
 \label{eq:S-jacobi-Be}\\
 \beta_e&=2\sum_{j=0}^{\min((s-e-1)/2,\rho_e)}\frac{\binom{s-e}{(s-e-1)/2-j}}{\binom{s-e}{(s-e-1)/2}}\frac{\Vol_{\rho_e-j}^{(Q)}(N+1)}{w_{\rho_e}^{(Q)}(N)}.
 \label{eq:S-jacobi-beta}
\end{align}
Here the imbalance between the numbers of $u$ and $v$ terms is $2j+1$, and the factor two accounts for its two orientations. If $s$ is odd, the same formula for $\beta_e$ with $e=0$ defines $\beta_0$; with $\rho_0=(r-1)/2$ the corresponding one-sided weight is
\begin{equation}
 B_0=\binom{s}{(s-1)/2}w_{\rho_0}^{(Q)}(N)=\frac{2D_1}{(Q-2)(s+1)}.
 \label{eq:S-jacobi-B0}
\end{equation}
Therefore
\begin{equation}
 C_s=\sum_{c\in\mathcal C_s}\alpha_cD_c+\sum_{e\in\mathcal E_s}\beta_eB_e+\mathbf 1_{\{s\ {\rm odd}\}}\beta_0B_0.
 \label{eq:S-jacobi-C-node-edge}
\end{equation}

For an ordinary edge $e$, the neighboring nodes are $e-1$ and $e+1$. Writing $\rho=\rho_e$, direct cancellation gives
\begin{align}
 \frac{B_e^2}{D_{e-1}D_{e+1}}&=\frac{\binom{s}{e}^2}{\binom{s}{e-1}\binom{s}{e+1}}\frac{\binom{s-e}{(s-e-1)/2}^2}{\binom{s-e+1}{(s-e+1)/2}\binom{s-e-1}{(s-e-1)/2}}\frac{w_{\rho}^{(Q)}(N)^2}{w_{\rho+1}^{(Q)}(N)w_{\rho}^{(Q)}(N)}\notag\\
 &=\frac{(e+1)(\rho_e+1)}{(Q-1)e(N-\rho_e)}=: \gamma_e.
 \label{eq:S-jacobi-gamma}
\end{align}
Order the nodes in $\mathcal C_s$ by increasing $c$; then $e-1$ and $e+1$ are consecutive nodes on this parity chain. With $v_c=\sqrt{D_c}$, define a symmetric tridiagonal matrix $J$ by
\begin{equation}
 J_{c,c}=\alpha_c,\qquad J_{e-1,e+1}=J_{e+1,e-1}=\frac12\beta_e\sqrt{\gamma_e}\quad(e\in\mathcal E_s).
 \label{eq:S-jacobi-J}
\end{equation}
When $s$ is odd, add the one-sided contribution to the first diagonal entry,
\begin{equation}
 J_{1,1}\leftarrow J_{1,1}+\frac{\beta_0B_0}{D_1}=J_{1,1}+\frac{2\beta_0}{(Q-2)(s+1)}.
 \label{eq:S-jacobi-one-sided-update}
\end{equation}
For an ordinary edge,
\begin{equation}
 2J_{e-1,e+1}v_{e-1}v_{e+1}=\beta_e\sqrt{\gamma_eD_{e-1}D_{e+1}}=\beta_eB_e.
 \label{eq:S-jacobi-edge-recovery}
\end{equation}
Together with \eqref{eq:S-jacobi-B0} and \eqref{eq:S-jacobi-one-sided-update}, Eqs.~\eqref{eq:S-jacobi-I-nodes} and \eqref{eq:S-jacobi-C-node-edge} become
\begin{equation}
 C_s=v^{\mathsf T}Jv,\qquad I_s=v^{\mathsf T}v,\qquad \frac{C_s}{I_s}\le\lambda_{\max}(J).
 \label{eq:S-jacobi-Rayleigh}
\end{equation}
Thus the Hamming-ball intersection ratio is reduced to the largest eigenvalue of a nonnegative symmetric tridiagonal matrix.

To bound the largest eigenvalue of $J$, define two half-line Jacobi matrices $\mathsf K_0$ and $\mathsf K_1$, indexed by $j\ge0$, with zero diagonal and off-diagonal entries
\begin{equation}
 b_j^{(0)}=\sqrt{\frac{2j+2}{2j+1}},\qquad b_j^{(1)}=\sqrt{\frac{2j+3}{2j+2}}.
 \label{eq:S-jacobi-model-b}
\end{equation}
For the even chain, put
\begin{equation}
 \phi_j=(2j+1)\sqrt{\binom{2j}{j}}\,2^{-3j/2},\qquad j\ge0.
 \label{eq:S-jacobi-even-ground}
\end{equation}
For $j\ge0$,
\begin{equation}
 b_j^{(0)}\frac{\phi_{j+1}}{\phi_j}=\frac{2j+3}{\sqrt2(2j+1)},
 \label{eq:S-jacobi-even-ratio-plus}
\end{equation}
and for $j\ge1$,
\begin{equation}
 b_{j-1}^{(0)}\frac{\phi_{j-1}}{\phi_j}=\frac{2\sqrt2\,j}{2j+1}.
 \label{eq:S-jacobi-even-ratio-minus}
\end{equation}
The two terms add to $3/\sqrt2$, while at $j=0$ Eq.~\eqref{eq:S-jacobi-even-ratio-plus} already equals $3/\sqrt2$. Hence
\begin{equation}
 b_{j-1}^{(0)}\phi_{j-1}+b_j^{(0)}\phi_{j+1}=\frac{3}{\sqrt2}\phi_j,
 \label{eq:S-jacobi-ground-eigen}
\end{equation}
with the first term omitted at $j=0$. Therefore, for every finitely supported real sequence $x_j=\phi_jy_j$,
\begin{equation}
 \frac{3}{\sqrt2}\sum_{j\ge0}x_j^2-2\sum_{j\ge0}b_j^{(0)}x_jx_{j+1}=\sum_{j\ge0}b_j^{(0)}\phi_j\phi_{j+1}(y_j-y_{j+1})^2\ge0.
 \label{eq:S-jacobi-ground-state}
\end{equation}
Thus every finite leading principal truncation of $\mathsf K_0$ has largest eigenvalue at most $3/\sqrt2$. Since a zero-diagonal Jacobi matrix is bipartite and hence has spectrum symmetric about zero, its operator norm is also at most $3/\sqrt2$.

For the odd chain, put
\begin{equation}
 \psi_j=(j+1)\sqrt{\frac{2^j}{(2j+1)\binom{2j}{j}}},\qquad j\ge0.
 \label{eq:S-jacobi-odd-ground}
\end{equation}
Then
\begin{equation}
 \frac{\psi_{j+1}}{\psi_j}=\frac{j+2}{\sqrt{(j+1)(2j+3)}}.
 \label{eq:S-jacobi-odd-ratio}
\end{equation}
Consequently,
\begin{equation}
 b_j^{(1)}\frac{\psi_{j+1}}{\psi_j}=\frac{j+2}{\sqrt2(j+1)},\qquad b_{j-1}^{(1)}\frac{\psi_{j-1}}{\psi_j}=\frac{2j+1}{\sqrt2(j+1)}\quad(j\ge1).
 \label{eq:S-jacobi-odd-eigen-check}
\end{equation}
For $j\ge1$ these terms add to $3/\sqrt2$, while at the left boundary
\begin{equation}
 b_0^{(1)}\psi_1+\frac{1}{\sqrt2}\psi_0=\frac{3}{\sqrt2}\psi_0.
 \label{eq:S-jacobi-odd-boundary-eigen}
\end{equation}
Thus
\begin{equation}
 b_{j-1}^{(1)}\psi_{j-1}+b_j^{(1)}\psi_{j+1}=\frac{3}{\sqrt2}\psi_j\qquad(j\ge1).
 \label{eq:S-jacobi-odd-eigen}
\end{equation}
For every finitely supported real sequence $x_j=\psi_jy_j$, the corresponding ground-state identity is
\begin{equation}
 \frac{3}{\sqrt2}\sum_{j\ge0}x_j^2-2\sum_{j\ge0}b_j^{(1)}x_jx_{j+1}-\frac{1}{\sqrt2}x_0^2=\sum_{j\ge0}b_j^{(1)}\psi_j\psi_{j+1}(y_j-y_{j+1})^2\ge0.
 \label{eq:S-jacobi-odd-ground-state}
\end{equation}
Hence every finite leading principal truncation satisfies the stronger boundary estimate
\begin{equation}
 \mathsf K_1+\frac{1}{\sqrt2}P_0\preceq\frac{3}{\sqrt2}I,
 \label{eq:S-jacobi-odd-boundary}
\end{equation}
where $P_0$ is the rank-one projection onto the first node. In particular, every finite truncation of $\mathsf K_1$ has largest eigenvalue, and therefore operator norm, at most $3/\sqrt2$.

The two parity estimates give a direct bound for the Hamming-ball intersection ratio. Let $\epsilon\in\{0,1\}$ be the parity of $s$, write the nodes as $c_j=2j+\epsilon$ in increasing order, and let $e_j=c_j+1$ denote the ordinary edge between $c_j$ and $c_{j+1}$. Define
\begin{equation}
 A_s=\max_{c\in\mathcal C_s}\alpha_c,
 \qquad
 \tau_s^{\rm ord}=\max_{e_j\in\mathcal E_s}\frac{\beta_{e_j}\sqrt{\gamma_{e_j}}}{2b_j^{(\epsilon)}},
 \label{eq:S-jacobi-majorant-scales}
\end{equation}
where the second maximum is taken to be zero if $\mathcal E_s$ is empty. When $s$ is even, set $\tau_s=\tau_s^{\rm ord}$. When $s$ is odd, the one-sided update in Eq.~\eqref{eq:S-jacobi-one-sided-update} has size
\begin{equation}
 \delta_s=\frac{2\beta_0}{(Q-2)(s+1)},
\end{equation}
and set
\begin{equation}
 \tau_s=\max\{\tau_s^{\rm ord},\sqrt2\,\delta_s\}.
 \label{eq:S-jacobi-tau}
\end{equation}
Let $\mathsf K_{\epsilon,s}$ be the leading principal truncation of $\mathsf K_\epsilon$ with $|\mathcal C_s|$ nodes. After reindexing $J$ by the node order $c_0,c_1,\ldots$, the definitions of $A_s$ and $\tau_s$ give the entrywise bounds
\begin{align}
 J&\le_{\rm ent} A_s I+\tau_s\mathsf K_{0,s}, && s\ \text{even},
 \label{eq:S-jacobi-dom-even}\\
 J&\le_{\rm ent} A_s I+\tau_s\left(\mathsf K_{1,s}+\frac{1}{\sqrt2}P_0\right), && s\ \text{odd}.
 \label{eq:S-jacobi-dom-odd}
\end{align}
All matrices in these entrywise comparisons are symmetric and nonnegative. Perron--Frobenius monotonicity therefore bounds the largest eigenvalue of $J$ by that of the corresponding majorant. Using Eq.~\eqref{eq:S-jacobi-ground-state} in the even case and Eq.~\eqref{eq:S-jacobi-odd-boundary} in the odd case yields
\begin{equation}
 \lambda_{\max}(J)\le A_s+\frac{3}{\sqrt2}\tau_s.
 \label{eq:S-jacobi-final-bound}
\end{equation}
Combining this with Eq.~\eqref{eq:S-jacobi-Rayleigh} gives the Jacobi bound
\begin{equation}
 \frac{C_s}{I_s}\le A_s+\frac{3}{\sqrt2}\tau_s.
 \label{eq:S-jacobi-ratio-bound}
\end{equation}

\smsection{High-alphabet half-gap at critical length}{sec:S3}

After Sec.~\ref{sec:S2}, it remains for $Q\ge16$ to treat the critical length $n=4t+1$. We first reduce separation propagation to an upper bound on the Hamming-ball intersection ratio, after which the Jacobi estimate of Sec.~\ref{sec:Sjacobi} will be applied. Put
\begin{equation}
 C_s=c_t^{(Q)}(4t+1,s),\qquad I_s=C_s-C_{s+1}.
 \label{eq:S-CsIs-critical}
\end{equation}
For $2\le s\le2t-1$, set
\begin{equation}
 x=s-1,\qquad L_s=\Kraw_t^{(Q)}(x;4t),\qquad M_s=\Kraw_{t-1}^{(Q)}(x;4t-1).
 \label{eq:S-LsMs-critical}
\end{equation}
The first-difference identity
\begin{equation}
 \Kraw_t^{(Q)}(x+1;N)-\Kraw_t^{(Q)}(x;N)=-Q\Kraw_{t-1}^{(Q)}(x;N-1)
 \label{eq:S-kraw-first-diff-critical}
\end{equation}
follows directly from the generating function. With $N=4t$ and $x=s-1$, it gives
\begin{equation}
 L_{s+1}=L_s-QM_s.
 \label{eq:S-L-difference-critical}
\end{equation}
For the interior propagation argument we henceforth assume $t\ge2$; when $t=1$, there is no integer $s$ with $2\le s\le2t-1$. We shall prove
\begin{equation}
 0<\frac{M_s}{L_s}\le\frac{t}{D_Q(t,s)},\qquad D_Q(t,s)=4(Q-1)t-\frac{9Q}{8}(s-1).
 \label{eq:S-L-ratio-critical}
\end{equation}
Once $L_s>0$ and $D_Q(t,s)>0$ are established below, the upper-bound part of Eq.~\eqref{eq:S-L-ratio-critical} is equivalent to the cleared inequality
\begin{equation}
 t\Kraw_t^{(Q)}(x;4t)-D_Q(t,x+1)\Kraw_{t-1}^{(Q)}(x;4t-1)\ge0,\qquad1\le x\le2t-2.
 \label{eq:S-contiguous-ratio-cleared}
\end{equation}
We prove the positivity and the ratio bound together by a continued-fraction comparison. Put
\begin{equation}
 P_j=\Kraw_j^{(Q)}(x;4t-1),\qquad r_j=\frac{P_{j-1}}{P_j}\quad(1\le j\le t).
 \label{eq:S-critical-Pj-rj}
\end{equation}
The length recursion gives
\begin{equation}
 L_s=P_t+(Q-1)P_{t-1},\qquad M_s=P_{t-1},\qquad \frac{M_s}{L_s}=\frac{r_t}{1+(Q-1)r_t}.
 \label{eq:S-critical-M-over-L-r}
\end{equation}
The fixed-length degree recurrence for the $Q$-ary Krawtchouk polynomials is
\begin{equation}
 jP_j=A_jP_{j-1}-B_jP_{j-2},\qquad A_j=(Q-1)(4t-j)+j-1-Qx,\qquad B_j=(Q-1)(4t-j+1).
 \label{eq:S-critical-degree-recurrence}
\end{equation}
Thus, whenever $r_{j-1}>0$,
\begin{equation}
 r_j=\frac{j}{A_j-B_jr_{j-1}}.
 \label{eq:S-critical-rj-recurrence}
\end{equation}
Set
\begin{equation}
 E_j=(Q-1)(4t-j)-Qx\left(1+\frac{j}{8t}\right),\qquad1\le j\le t.
 \label{eq:S-critical-Ej}
\end{equation}
For $Q\ge16$, $t\ge2$, and $1\le x\le2t-2$, these quantities are positive, since $E_j$ is minimized at $j=t$, $x=2t-2$, where
\begin{equation}
 E_t\ge\left(\frac{3Q}{4}-3\right)t+\frac{9Q}{4}>0.
 \label{eq:S-critical-Ej-positive}
\end{equation}
We claim that
\begin{equation}
 0<r_j\le\frac{j}{E_j}\qquad(1\le j\le t).
 \label{eq:S-critical-rj-barrier}
\end{equation}
For $j=1$, one has $r_1=1/A_1$ and $A_1>E_1>0$. Suppose the claim holds at $j-1$. Then the denominator in Eq.~\eqref{eq:S-critical-rj-recurrence} is at least $A_j-B_j(j-1)/E_{j-1}$, so it is enough to verify
\begin{equation}
 A_j-\frac{B_j(j-1)}{E_{j-1}}\ge E_j.
 \label{eq:S-critical-barrier-step}
\end{equation}
After multiplication by the positive quantity $E_{j-1}$, the left-hand side minus the right-hand side equals
\begin{equation}
 \frac{Qx}{64t^2}\,H_j(x),
 \label{eq:S-critical-barrier-residual}
\end{equation}
where
\begin{align}
 H_j(x)=&-8Qj^2t-Qj^2x+32Qjt^2-8Qjtx+8Qjt+Qjx\notag\\
 &{}-96jt^2+8jt+64t^2-8t.
 \label{eq:S-critical-Hj}
\end{align}
This polynomial is decreasing in $x$. At $x=2t-2$, the quantity $H_j(2t-2)/2$, viewed as a polynomial in $j$, is concave on $[2,t]$. Hence its minimum on the integer interval $2\le j\le t$ occurs at an endpoint, where
\begin{align}
 \frac{H_2(2t-2)}{2}&=2\{8Qt^2+3Qt+Q-32t^2+2t\}>0,\notag\\
 \frac{H_t(2t-2)}{2}&=t\{(3Q-48)t^2+(14Q+36)t-(Q+4)\}>0.
 \label{eq:S-critical-H-endpoints}
\end{align}
The inequalities use $Q\ge16$ and $t\ge2$. Thus Eq.~\eqref{eq:S-critical-barrier-step} holds, and the induction proves Eq.~\eqref{eq:S-critical-rj-barrier}. Since $P_0=1$, the positivity of the ratios also gives $P_j>0$ for $0\le j\le t$. In particular, $M_s=P_{t-1}>0$ and $L_s=P_t+(Q-1)P_{t-1}>0$. Taking $j=t$ in Eq.~\eqref{eq:S-critical-rj-barrier} gives
\begin{equation}
 r_t\le\frac{t}{3(Q-1)t-\frac{9Q}{8}x}.
 \label{eq:S-critical-rt-bound}
\end{equation}
Since $r\mapsto r/(1+(Q-1)r)$ is increasing for $r\ge0$, Eqs.~\eqref{eq:S-critical-M-over-L-r} and \eqref{eq:S-critical-rt-bound} give
\begin{equation}
 \frac{M_s}{L_s}\le\frac{t}{3(Q-1)t-\frac{9Q}{8}x+(Q-1)t}=\frac{t}{D_Q(t,x+1)}.
 \label{eq:S-critical-ratio-from-barrier}
\end{equation}
Together with positivity, this proves Eq.~\eqref{eq:S-L-ratio-critical}. Moreover,
\begin{equation}
 \frac{D_Q(t,s)}{t}\ge4(Q-1)-\frac{9Q}{8}\frac{2t-2}{t}>Q,
 \label{eq:S-DQ-positive-critical}
\end{equation}
so $0<Qt/D_Q(t,s)<1$. Writing $u_s=M_s/L_s$, we therefore have
\begin{equation}
 \frac{R_{Q;4t+1,t}(s+1)}{R_{Q;4t+1,t}(s)}=\frac{1-I_s/C_s}{(1-Qu_s)^2}.
 \label{eq:S-R-step-factor-critical}
\end{equation}
By Eq.~\eqref{eq:S-jacobi-I-nodes}, $I_s>0$. Hence it is enough to prove
\begin{equation}
 \frac{C_s}{I_s}\le T_Q(t,s),\qquad T_Q(t,s)=\frac{D_Q(t,s)^2}{Qt\{2D_Q(t,s)-Qt\}}.
 \label{eq:S-target-critical}
\end{equation}
Indeed, Eq.~\eqref{eq:S-target-critical} is equivalent to
\begin{equation}
 \frac{I_s}{C_s}\ge\frac{Qt}{D_Q(t,s)}\left(2-\frac{Qt}{D_Q(t,s)}\right).
 \label{eq:S-target-IoverC-critical}
\end{equation}
The right-hand side is positive by Eq.~\eqref{eq:S-DQ-positive-critical}. Since $u_s\le t/D_Q(t,s)$ and $v\mapsto Qv(2-Qv)$ is increasing on $0\le v\le t/D_Q(t,s)<1/Q$, Eq.~\eqref{eq:S-target-IoverC-critical} implies $I_s/C_s\ge Qu_s(2-Qu_s)$, which is exactly Eq.~\eqref{eq:S-R-step-factor-critical}$\le1$. For $Q=16$, with $r=2t-s$,
\begin{equation}
 T_{16}(t,s)=\frac{9(4t+3r+3)^2}{16t(8t+9r+9)}.
 \label{eq:S-T16-critical}
\end{equation}

Put
\begin{equation}
 r=2t-s,\qquad N=4t-s.
 \label{eq:S-rN-critical}
\end{equation}
At the critical length, write $A_s^{(Q)}$ and $\tau_s^{(Q)}$ for the quantities $A_s$ and $\tau_s$ of Sec.~\ref{sec:Sjacobi}. Equation~\eqref{eq:S-jacobi-ratio-bound} gives
\begin{equation}
 \frac{C_s}{I_s}\le A_s^{(Q)}+\frac{3}{\sqrt2}\tau_s^{(Q)}.
 \label{eq:S-critical-jacobi-bound}
\end{equation}
Thus Eq.~\eqref{eq:S-target-critical} follows once
\begin{equation}
 A_s^{(Q)}+\frac{3}{\sqrt2}\tau_s^{(Q)}\le T_Q(t,s).
 \label{eq:S-J-target-critical}
\end{equation}

We first reduce the alphabet to $Q=16$. For $0\le a\le b\le N$, the ball--shell quotient appearing in Eqs.~\eqref{eq:S-jacobi-alpha} and \eqref{eq:S-jacobi-beta} has the expansion
\[
 \frac{\Vol_a^{(Q)}(N+1)}{w_b^{(Q)}(N)}=\frac{1}{\binom Nb}\sum_{j=0}^a\binom{N+1}{j}(Q-1)^{j-b},
\]
so it is nonincreasing in $Q$. Hence every $\alpha_c$ and $\beta_e$, including $\beta_0$, is nonincreasing in $Q$. Equation~\eqref{eq:S-jacobi-gamma} shows that each $\gamma_e$ is also decreasing, while the odd one-sided factor $(Q-2)^{-1}$ decreases. Therefore the scales in Eqs.~\eqref{eq:S-jacobi-majorant-scales} and \eqref{eq:S-jacobi-tau} satisfy
\begin{equation}
 A_s^{(Q)}\le A_s^{(16)},\qquad \tau_s^{(Q)}\le\tau_s^{(16)}\qquad(Q\ge16).
 \label{eq:S-J-monotone-critical}
\end{equation}
On the target side, write
\begin{equation}
 z_Q=\frac{D_Q(t,s)}{Qt}=4-\frac4Q-\frac98\frac{s-1}{t}.
 \label{eq:S-zQ-critical}
\end{equation}
Then $z_Q$ is increasing in $Q$, and $z_{16}>1$ for $2\le s\le2t-1$. Since $g(z)=z^2/(2z-1)$ is increasing for $z>1$,
\begin{equation}
 T_Q(t,s)=g(z_Q)\ge g(z_{16})=T_{16}(t,s).
 \label{eq:S-target-monotone-critical}
\end{equation}
For the nonterminal shells it is therefore enough to prove Eq.~\eqref{eq:S-J-target-critical} at $Q=16$.

Set
\begin{equation}
 \zeta_*=\frac{r}{15(4t+r+2)},\qquad A(\zeta)=\frac{(1+15\zeta)(1+\zeta)}{(1-\zeta)^2},\qquad B(\zeta)=\frac{2(1+15\zeta)}{(1-\zeta)^2}.
 \label{eq:S-AB-zeta-critical}
\end{equation}
For a node $c\in\mathcal C_s$, put $\zeta_c=\ell_c/[15(N-\ell_c+1)]$. Then $0\le\zeta_c\le\zeta_*$, and the shell-tail estimate \eqref{eq:S-shell-tail} gives, for $0\le j\le\ell_c$,
\begin{equation}
 \frac{\Vol_{\ell_c-j}^{(16)}(N+1)}{w_{\ell_c}^{(16)}(N)}\le(1+15\zeta_c)\frac{\zeta_c^j}{1-\zeta_c}.
 \label{eq:S-node-tail-to-zeta}
\end{equation}
The central-binomial ratios in Eq.~\eqref{eq:S-jacobi-alpha} are at most one, so summing the geometric tail gives
\begin{equation}
 \alpha_c^{(16)}\le A(\zeta_c)\le A(\zeta_*).
 \label{eq:S-alpha-sum-to-A}
\end{equation}
For an ordinary edge $e=c+1$, one has $\rho_e+1=\ell_c$, hence the corresponding shell parameter $(\rho_e+1)/[15(N-\rho_e)]$ equals $\zeta_c$. The same argument applied to Eq.~\eqref{eq:S-jacobi-beta} gives
\begin{equation}
 \beta_e^{(16)}\le B(\zeta_c)\le B(\zeta_*).
 \label{eq:S-beta-sum-to-B}
\end{equation}
If $s$ is odd, the one-sided coefficient has the sharper parameter $\widetilde\zeta_0=(r-1)/[15(4t+r+3)]\le\zeta_*$, and therefore $\beta_0^{(16)}\le B(\zeta_*)$.

Write the nodes as $c_j=2j+\epsilon$, where $\epsilon$ is the parity of $s$. For the ordinary edge $e_j=c_j+1$, Eq.~\eqref{eq:S-jacobi-gamma} becomes
\begin{equation}
 \gamma_{e_j}^{(16)}=(b_j^{(\epsilon)})^2\zeta_{c_j}\le(b_j^{(\epsilon)})^2\zeta_*.
 \label{eq:S-gamma-model-critical}
\end{equation}
Equations~\eqref{eq:S-jacobi-majorant-scales}, \eqref{eq:S-beta-sum-to-B}, and \eqref{eq:S-gamma-model-critical} therefore give
\begin{equation}
 (\tau_s^{\rm ord})^{(16)}\le\frac12B(\zeta_*)\sqrt{\zeta_*}.
 \label{eq:S-critical-tau-ordinary}
\end{equation}
When $s$ is odd, the one-sided diagonal load in Sec.~\ref{sec:Sjacobi} satisfies
\begin{equation}
 \delta_s^{(16)}=\frac{\beta_0^{(16)}}{7(s+1)}\le\frac{B(\zeta_*)}{7(s+1)}.
 \label{eq:S-special-d-critical}
\end{equation}
For a nonterminal odd chain $r\ge3$, both $r$ and $s$ are odd and at least three. Since $t=(r+s)/2$,
\begin{equation}
 \zeta_*=\frac{r}{15(3r+2s+2)}.
 \label{eq:S-zeta-rs-critical}
\end{equation}
Moreover,
\begin{equation}
 49r(s+1)^2-120(3r+2s+2)\ge147(s+1)^2-240(s+1)-1080>0,
 \label{eq:S-special-absorb-poly}
\end{equation}
which, together with Eq.~\eqref{eq:S-special-d-critical}, implies
\begin{equation}
 \sqrt2\,\delta_s^{(16)}\le\frac12B(\zeta_*)\sqrt{\zeta_*}.
 \label{eq:S-special-absorb-critical}
\end{equation}
Thus Eq.~\eqref{eq:S-jacobi-tau} and the even definition of $\tau_s$ give, for every nonterminal shell,
\begin{equation}
 A_s^{(16)}\le A(\zeta_*),\qquad \tau_s^{(16)}\le\frac12B(\zeta_*)\sqrt{\zeta_*}.
 \label{eq:S-critical-scales}
\end{equation}
Combining this with Eq.~\eqref{eq:S-critical-jacobi-bound} yields
\begin{equation}
 \frac{C_s}{I_s}\le A(\zeta_*)+\frac{3}{2\sqrt2}B(\zeta_*)\sqrt{\zeta_*}.
 \label{eq:S-perron-even-critical}
\end{equation}
The terminal odd case $r=1$ is one-dimensional. Here $s=2t-1$, and the formulas of Sec.~\ref{sec:Sjacobi} give, for every $Q\ge16$,
\begin{equation}
 \frac{C_s}{I_s}=1+\frac{2}{(Q-2)t}\le1+\frac{1}{7t}<T_{16}(t,2t-1)\le T_Q(t,2t-1),
 \label{eq:S-terminal-r1-critical}
\end{equation}
where
\[
 T_{16}(t,2t-1)-\left(1+\frac{1}{7t}\right)=\frac{28t^2+220t+495}{56t(4t+9)}>0.
\]

It remains to compare Eq.~\eqref{eq:S-perron-even-critical} with the target. Put
\begin{equation}
 u=\frac{r}{t},\qquad \bar\zeta=\frac{u}{15(4+u)}.
 \label{eq:S-zeta-bar-critical}
\end{equation}
Then $0\le\zeta_*\le\bar\zeta\le1/45$, and $A$ and $B$ are increasing on this interval. Moreover,
\begin{equation}
 \Theta(\zeta)=\frac{9(1+30\zeta)^2}{(1-15\zeta)(8+420\zeta)},\qquad T_{16}(t,s)\ge\Theta(\bar\zeta).
 \label{eq:S-Theta-critical}
\end{equation}
Indeed, $\Theta(\bar\zeta)=9(3u+4)^2/[16(9u+8)]$, and direct subtraction gives
\[
 T_{16}(t,s)-\Theta(\bar\zeta)=\frac{27\{27u^2+27u/t+48u+24/t+16\}}{16t(9u+8)(9u+9/t+8)}>0.
\]
For $0\le\zeta\le1/45$,
\begin{equation}
 \Theta(\zeta)\ge A(\zeta)+\frac{3}{2\sqrt2}B(\zeta)\sqrt\zeta.
 \label{eq:S-scalar-critical}
\end{equation}
Indeed,
\begin{equation}
 \frac{3}{2\sqrt2}\sqrt\zeta\le\frac{15}{2}\zeta+\frac{3}{80},\qquad \frac{15}{2}\zeta+\frac{3}{80}-\frac{3}{2\sqrt2}\sqrt\zeta=\frac{3}{80}(10\sqrt{2\zeta}-1)^2,
 \label{eq:S-root-linear-critical}
\end{equation}
and direct reduction gives
\begin{equation}
 \Theta(\zeta)-A(\zeta)-B(\zeta)\left(\frac{15}{2}\zeta+\frac{3}{80}\right)=\frac{15201000\zeta^4+1147275\zeta^3+22440\zeta^2-575\zeta+4}{40(1-\zeta)^2(1-15\zeta)(105\zeta+2)}.
 \label{eq:S-scalar-rational-critical}
\end{equation}
The denominator is positive on $[0,1/45]$. The numerator is
\begin{equation}
 \zeta^3(15201000\zeta+1147275)+(22440\zeta^2-575\zeta+4),
 \label{eq:S-scalar-numerator-critical}
\end{equation}
and the final quadratic has discriminant
\begin{equation}
 575^2-4\cdot22440\cdot4=-28415<0.
 \label{eq:S-scalar-discriminant-critical}
\end{equation}
Therefore Eq.~\eqref{eq:S-J-target-critical} holds at $Q=16$ for every nonterminal shell. Together with Eqs.~\eqref{eq:S-J-monotone-critical} and \eqref{eq:S-target-monotone-critical}, this proves Eq.~\eqref{eq:S-target-critical} for every nonterminal shell and every $Q\ge16$; the terminal shell was handled directly in Eq.~\eqref{eq:S-terminal-r1-critical}.

The initial separations are independent of the interior Jacobi comparison. By the Lloyd shift identity, at $n=4t+1$,
\begin{equation}
 \Lloyd_t^{(Q,4t+1)}(s)=\Kraw_t^{(Q)}(s-1;4t).
 \label{eq:S-critical-L-shift}
\end{equation}
Hence
\begin{equation}
 \Lloyd_t^{(Q,4t+1)}(1)=(Q-1)^t\binom{4t}{t},
 \label{eq:S-L-s1}
\end{equation}
and
\begin{align}
 \Lloyd_t^{(Q,4t+1)}(2)&=(Q-1)^t\binom{4t-1}{t}-(Q-1)^{t-1}\binom{4t-1}{t-1}\notag\\
 &=(Q-1)^{t-1}\binom{4t-1}{t}\frac{3Q-4}{3}.
 \label{eq:S-L-s2}
\end{align}
The separation-one intersection is
\begin{equation}
 c_t^{(Q)}(4t+1,1)=Q\Vol_{t-1}^{(Q)}(4t).
 \label{eq:S-c-s1}
\end{equation}
For $t\ge2$, the separation-two intersection is
\begin{equation}
 c_t^{(Q)}(4t+1,2)=2\Vol_{t-1}^{(Q)}(4t-1)+(Q^2-2)\Vol_{t-2}^{(Q)}(4t-1).
 \label{eq:S-c-s2}
\end{equation}
When $t=1$, the same local classification gives directly $c_1^{(Q)}(5,2)=2$.

Put $\eta=1/[3(Q-1)]$. For separation one, Eq.~\eqref{eq:S-shell-tail} at length $4t$ gives $\Vol_{t-1}^{(Q)}(4t)\le\eta\Vol_t^{(Q)}(4t)$. Since $w_t^{(Q)}(4t)=\Vol_t^{(Q)}(4t)-\Vol_{t-1}^{(Q)}(4t)$,
\[
 \Vol_t^{(Q)}(4t)\le\frac{w_t^{(Q)}(4t)}{1-\eta},\qquad \Vol_{t-1}^{(Q)}(4t)\le\frac{\eta\,w_t^{(Q)}(4t)}{1-\eta}.
\]
Together with $\Vol_t^{(Q)}(4t+1)=\Vol_t^{(Q)}(4t)+(Q-1)\Vol_{t-1}^{(Q)}(4t)$ and Eq.~\eqref{eq:S-L-s1}, this yields
\begin{equation}
 R_{Q;4t+1,t}(1)\le\frac{4Q(Q-1)}{(3Q-4)^2}.
 \label{eq:S-end1}
\end{equation}

For separation two and $t\ge2$, set $M=4t-1$. Equation~\eqref{eq:S-shell-tail}, first with $j=t$ and then with $j=t-1$, gives
\[
 \Vol_{t-1}^{(Q)}(M)\le\frac{\eta\,w_t^{(Q)}(M)}{1-\eta},\qquad \Vol_{t-2}^{(Q)}(M)\le\frac{\eta^2w_t^{(Q)}(M)}{1-\eta}.
\]
Applying the length recursion twice,
\[
 \Vol_t^{(Q)}(M+2)\le\frac{(1+(Q-1)\eta)^2}{1-\eta}w_t^{(Q)}(M),
\]
while Eq.~\eqref{eq:S-c-s2} gives
\[
 c_t^{(Q)}(4t+1,2)\le\frac{2\eta+(Q^2-2)\eta^2}{1-\eta}w_t^{(Q)}(M).
\]
Since Eq.~\eqref{eq:S-L-s2} is equivalently $\Lloyd_t^{(Q,4t+1)}(2)=w_t^{(Q)}(M)(3Q-4)/[3(Q-1)]$, we obtain
\begin{equation}
 R_{Q;4t+1,t}(2)\le\frac{16(Q-1)^2(Q^2+6Q-8)}{(3Q-4)^4}\qquad(t\ge2).
 \label{eq:S-end2}
\end{equation}
The two endpoint bounds satisfy
\begin{equation}
 \frac12-\frac{4Q(Q-1)}{(3Q-4)^2}=\frac{Q^2-16Q+16}{2(3Q-4)^2}\ge0,
 \label{eq:S-end1-check}
\end{equation}
and
\begin{equation}
 \frac12-\frac{16(Q-1)^2(Q^2+6Q-8)}{(3Q-4)^4}=\frac{(7Q-8)(7Q^3-72Q^2+128Q-64)}{2(3Q-4)^4}>0.
 \label{eq:S-end2-check}
\end{equation}
For $Q\ge16$, the denominator in each expression is positive; in Eq.~\eqref{eq:S-end1-check} the numerator is increasing for $Q\ge8$ and positive at $Q=16$, while in Eq.~\eqref{eq:S-end2-check} both factors are positive and the cubic factor is increasing for $Q\ge16$. For the remaining case $t=1$, one has
\begin{equation}
 R_{Q;5,1}(2)=\frac{2(5Q-4)}{(3Q-4)^2},\qquad \frac12-R_{Q;5,1}(2)=\frac{(Q-4)(9Q-8)}{2(3Q-4)^2}>0.
 \label{eq:S-end2-t1}
\end{equation}
Thus $R_{Q;4t+1,t}(1)\le1/2$ for every $t\ge1$, while $R_{Q;4t+1,t}(2)<1/2$ for every $t\ge1$.

\begin{theorem}[Uniform high-alphabet half-gap]
\label{thm:S-halfgap}
For every integer $t\ge1$, every $Q\ge16$, every $n\ge4t+1$, and every $1\le s\le2t$,
\begin{equation}
 \Lloyd_t^{(Q,n)}(s)>0,\qquad R_{Q;n,t}(s)\le\frac12.
\end{equation}
\end{theorem}

\begin{proof}
Section~\ref{sec:S2} gives $\Lloyd_t^{(Q,n)}(s)>0$ for $n\ge4t+1$ and reduces the ratio bound to the critical length $n=4t+1$. If $t=1$, the two possible separations are exactly the endpoint cases above. Let $t\ge2$. The target reduction in Eq.~\eqref{eq:S-target-critical} and the estimates of the preceding block give
\[
 R_{Q;4t+1,t}(s+1)\le R_{Q;4t+1,t}(s)\qquad(2\le s\le2t-1).
\]
Hence $R_{Q;4t+1,t}(s)\le R_{Q;4t+1,t}(2)<1/2$ for $2\le s\le2t$, while Eq.~\eqref{eq:S-end1-check} gives $R_{Q;4t+1,t}(1)\le1/2$. The length monotonicity of Sec.~\ref{sec:S2} then gives the same ratio bound for every $n\ge4t+1$.
\end{proof}

\smsection{Qutrit exterior ranges}{sec:S4}

This section disposes of all qutrit lengths outside the bridge. The short range is handled by the Singleton bound and a volume estimate; the long range uses a positive-Lloyd Jacobi comparison. Set $Q=9$ and abbreviate
\begin{equation}
 \Vol_t(n)=\sum_{j=0}^{t}8^j\binom nj.
\end{equation}
The length line is divided into
\begin{align}
&17n\le72t,
\label{eq:S-short}\\
&84(n-1)\ge373t,
\label{eq:S-long}\\
&\frac{72}{17}t<n<1+\frac{373}{84}t.
\label{eq:S-bridge}
\end{align}
If the first two alternatives fail, the remaining integer triples are exactly in the bridge window \eqref{eq:S-bridge}.

In the short range, the quantum Singleton bound gives $K\le3^{n-4t}$ for nontrivial qutrit codes. It remains to prove
\begin{equation}
 \Vol_t(n)\le3^{4t}\qquad(17n\le72t).
 \label{eq:S-qutrit-vol}
\end{equation}
Put $\alpha=t/n$. In the nontrivial range Singleton leaves only $n\ge4t$, so here $17/72\le\alpha\le1/4$. For every $0<\lambda\le1$,
\begin{equation}
 \sum_{j=0}^{t}8^j\binom nj\le\lambda^{-t}(1+8\lambda)^n,
\end{equation}
because $\lambda^{-t}(1+8\lambda)^n=\sum_{j=0}^{n}8^j\binom nj\lambda^{j-t}$ and $\lambda^{j-t}\ge1$ for $j\le t$. Taking $\lambda=t/[8(n-t)]\le1$ gives
\begin{equation}
 \sum_{j=0}^{t}8^j\binom nj\le\frac{n^n8^t}{t^t(n-t)^{n-t}}.
 \label{eq:S-entropy-bound}
\end{equation}
Thus Eq.~\eqref{eq:S-qutrit-vol} is reduced to
\begin{equation}
 \frac{8^\alpha}{\alpha^\alpha(1-\alpha)^{1-\alpha}}\le81^\alpha.
 \label{eq:S-alpha-bound}
\end{equation}
Let
\begin{equation}
 \Phi(\alpha)=-\alpha\log\alpha-(1-\alpha)\log(1-\alpha)+\alpha\log8-\alpha\log81.
\end{equation}
Then
\begin{equation}
 \Phi'(\alpha)=\log\frac{8(1-\alpha)}{81\alpha}<0\qquad(\alpha\ge17/72),
\end{equation}
so the left endpoint suffices. Its exact integer form is
\begin{equation}
 8^{17}72^{72}<81^{17}55^{55}17^{17},
 \label{eq:S-entropy-end}
\end{equation}
which is exactly $\Phi(17/72)<0$.

The long range is covered by the positive-Lloyd Jacobi estimate. It is convenient to use the slightly stronger natural threshold encoded by
\begin{equation}
 c_0=\frac{27+6\sqrt2}{8},\qquad h=1+3\sqrt2.
 \label{eq:S-qutrit-c0h}
\end{equation}
These constants satisfy
\begin{equation}
 8c_0-25=2h,\qquad8(c_0-1)=h^2,\qquad\frac{373}{84}>c_0.
 \label{eq:S-qutrit-c0-identities}
\end{equation}
Thus the present long range $n-1\ge(373/84)t$ lies inside the natural positive-Lloyd range $n-1\ge c_0t$.

First, the Lloyd response is positive throughout the overlap range. Let $\xi_1$ be the first zero of $P_t^{(9)}(x;n-1)$. The Jacobi-matrix estimate of Sec.~\ref{sec:S2}, now with $Q=9$, gives
\begin{equation}
 \xi_1\ge\frac{8N-7(t-1)-2\sqrt{8(t-1)(N-t+2)}}{9},\qquad N=n-1.
 \label{eq:S-qutrit-natural-zero}
\end{equation}
For $N\ge c_0t$, the left-hand side of
\begin{equation}
 8N-25t+16>2\sqrt{8(t-1)(N-t+2)}
 \label{eq:S-qutrit-natural-zero-step}
\end{equation}
is positive. After squaring, the difference between the two sides squared is increasing in $N$, and at $N=c_0t$ it equals
\begin{equation}
 4\bigl((19+54\sqrt2)t+80\bigr)>0.
 \label{eq:S-qutrit-natural-zero-end}
\end{equation}
Hence $\xi_1>2t-1$. By the Lloyd shift identity of Sec.~\ref{sec:S1},
\begin{equation}
 \Lloyd_t^{(9,n)}(s)=P_t^{(9)}(s-1;n-1).
\end{equation}
Since $s-1\le2t-1<\xi_1$ and $P_t^{(9)}(0;n-1)=8^t\binom{n-1}{t}>0$, no zero is crossed and therefore $\Lloyd_t^{(9,n)}(s)>0$ for $1\le s\le2t$.

It remains to prove the raw Lloyd-square comparison. For the interior separation step assume $t\ge2$; when $t=1$, there is no integer $s$ with $2\le s\le2t-1$. For $2\le s\le2t-1$ put
\begin{equation}
 r=2t-s,\qquad u=\frac{r}{t},\qquad\kappa(u)=\frac{1}{h+\frac92u}.
 \label{eq:S-qutrit-u-kappa}
\end{equation}
Let
\begin{equation}
 x=s-1,\qquad L_s=P_t^{(9)}(x;n-1),\qquad M_s=P_{t-1}^{(9)}(x;n-2).
 \label{eq:S-qutrit-LM}
\end{equation}
We first bound the contiguous ratio entering the Lloyd first difference. Put
\begin{equation}
 y=s-2,\qquad D=n-3,\qquad P_j=P_j^{(9)}(y;D).
 \label{eq:S-qutrit-riccati-notation}
\end{equation}
The degree recurrence for qutrit Krawtchouk polynomials is
\begin{equation}
 jP_j=\bigl(8D-7(j-1)-9y\bigr)P_{j-1}-8(D-j+2)P_{j-2},\qquad j\ge1,\qquad P_{-1}=0.
 \label{eq:S-qutrit-degree-recurrence}
\end{equation}
For $1\le j\le t-1$, set
\begin{equation}
 E_j=\kappa(u)\bigl(8D-7(j-1)-9y\bigr)-8\kappa(u)^2(D-j+2)-j.
 \label{eq:S-qutrit-riccati-residual}
\end{equation}
The residual is increasing in $D$ and decreasing in $j$, since
\begin{equation}
 \partial_D E_j=8\kappa(1-\kappa)>0,\qquad E_{j+1}-E_j=(\kappa-1)(8\kappa+1)<0.
\end{equation}
Since $D=n-3\ge c_0t-2$, the minimum over the required range occurs at $D=c_0t-2$ and $j=t-1$, where it equals
\begin{align}
&t\bigl[\kappa(u)(2h+9u)-h^2\kappa(u)^2-1\bigr]+16\kappa(u)-8\kappa(u)^2+1\notag\\
&=\frac{9t\,u(9u+4+12\sqrt2)}{(9u+2+6\sqrt2)^2}+\frac{27(3u^2+(12+4\sqrt2)u+4+8\sqrt2)}{(9u+2+6\sqrt2)^2}>0.
 \label{eq:S-qutrit-riccati-positive}
\end{align}
Hence $E_j>0$ throughout the required range. We now prove inductively that
\begin{equation}
 P_j>0,\qquad0<\frac{P_{j-1}}{P_j}\le\kappa(u),\qquad1\le j\le t-1.
 \label{eq:S-qutrit-riccati-bound}
\end{equation}
For $j=1$, $P_1=(8D-9y)P_0$ and $E_1>0$ imply $\kappa(u)(8D-9y)>1$, so $P_1>0$ and $P_0/P_1<\kappa(u)$. Suppose the claim holds through $j-1$. Dividing Eq.~\eqref{eq:S-qutrit-degree-recurrence} by $P_{j-1}>0$ gives
\begin{equation}
 \frac{jP_j}{P_{j-1}}=8D-7(j-1)-9y-8(D-j+2)\frac{P_{j-2}}{P_{j-1}}\ge\frac{j}{\kappa(u)}>0,
\end{equation}
where the inequality uses $P_{j-2}/P_{j-1}\le\kappa(u)$ and $E_j\ge0$. Thus $P_j>0$ and $P_{j-1}/P_j\le\kappa(u)$, proving Eq.~\eqref{eq:S-qutrit-riccati-bound}.

The two length/position shifts from the generating function are
\begin{align}
 P_{t-1}^{(9)}(y+1;D+1)&=P_{t-1}-P_{t-2},
 \label{eq:S-qutrit-shift-M}\\
 P_t^{(9)}(y+1;D+2)&=P_t+7P_{t-1}-8P_{t-2}.
 \label{eq:S-qutrit-shift-L}
\end{align}
Thus, with $\rho=P_{t-2}/P_{t-1}$,
\begin{equation}
 M_s=P_{t-1}(1-\rho),\qquad L_s=P_t+7P_{t-1}-8P_{t-2}.
\end{equation}
Equation~\eqref{eq:S-qutrit-riccati-bound} gives $0<\rho\le\kappa(u)<1$, so $M_s>0$; the Lloyd positivity proved above gives $L_s>0$. Using Eq.~\eqref{eq:S-qutrit-degree-recurrence} at degree $t$ gives
\begin{equation}
 \frac{L_s}{M_s}=\frac{8(n-1)-9(s-1)/(1-\rho)}{t}.
\end{equation}
The function $\rho\mapsto8(n-1)-9(s-1)/(1-\rho)$ is decreasing on $0\le\rho<1$, and therefore
\begin{equation}
 \frac{L_s}{M_s}\ge\frac{D_n(t,s)}{t},\qquad D_n(t,s)=8(n-1)-\frac{9(s-1)}{1-\kappa(u)}.
\end{equation}
The long-range hypothesis, $s-1<2t$, and $\kappa(u)\le1/h$ give
\begin{equation}
 \frac{D_n(t,s)}{t}>\frac{746}{21}-\frac{18}{1-1/h}=\frac{368}{21}-3\sqrt2>9.
 \label{eq:S-qutrit-dgt9}
\end{equation}
Thus $D_n(t,s)>0$, and taking reciprocals yields
\begin{equation}
 0<\frac{M_s}{L_s}\le\frac{t}{D_n(t,s)}.
 \label{eq:S-qutrit-Dn}
\end{equation}
Let $C_s=c_t^{(9)}(n,s)$ and $I_s=C_s-C_{s+1}$. By Eq.~\eqref{eq:S-jacobi-I-nodes}, $I_s>0$. Since $L_{s+1}=L_s-9M_s$, put $\Delta_s=9M_s/L_s$. The Lloyd positivity at $s$ and $s+1$, together with $M_s>0$, gives $0<\Delta_s<1$, and
\begin{equation}
 \frac{R_{9;n,t}(s+1)}{R_{9;n,t}(s)}=\frac{1-I_s/C_s}{(1-\Delta_s)^2}.
\end{equation}
Hence $R_{9;n,t}(s+1)\le R_{9;n,t}(s)$ follows from
\begin{equation}
 \frac{I_s}{C_s}\ge2\Delta_s-\Delta_s^2.
\end{equation}
By Eqs.~\eqref{eq:S-qutrit-Dn} and \eqref{eq:S-qutrit-dgt9}, $0<\Delta_s\le9t/D_n(t,s)<1$. Since $v\mapsto2v-v^2$ is increasing on $[0,1]$, it is enough to prove
\begin{equation}
 \frac{C_s}{I_s}\le T_n(t,s),\qquad T_n(t,s)=\frac{D_n(t,s)^2}{9t(2D_n(t,s)-9t)}.
 \label{eq:S-qutrit-target}
\end{equation}

We now apply the Jacobi representation of Sec.~\ref{sec:Sjacobi}. Set $N=n-s-1$ and
\begin{equation}
 q_*:=\frac{r}{8(2n-4t+r)}.
 \label{eq:S-qutrit-qstar}
\end{equation}
For a node $c\in\mathcal C_s$, put $q_c=\ell_c/[8(N-\ell_c+1)]$. Then
\begin{equation}
 q_c=\frac{r-c}{8(2n-4t+r+c)}\le q_*.
\end{equation}
The shell-tail estimate \eqref{eq:S-shell-tail} gives, for $0\le j\le\ell_c$,
\begin{equation}
 \frac{\Vol_{\ell_c-j}^{(9)}(N+1)}{w_{\ell_c}^{(9)}(N)}\le(1+8q_c)\frac{q_c^j}{1-q_c}.
 \label{eq:S-qutrit-node-tail}
\end{equation}
The central-binomial ratios in Eq.~\eqref{eq:S-jacobi-alpha} are at most one, so
\begin{equation}
 \alpha_c\le A(q_c)\le A(q_*),\qquad A(q)=\frac{(1+8q)(1+q)}{(1-q)^2}.
 \label{eq:S-qutrit-alpha-majorant}
\end{equation}
For an ordinary edge $e=c+1$, one has $\rho_e+1=\ell_c$, hence the shell parameter in Eq.~\eqref{eq:S-jacobi-beta} is the same $q_c$. Therefore
\begin{equation}
 \beta_e\le B(q_c)\le B(q_*),\qquad B(q)=\frac{2(1+8q)}{(1-q)^2}.
 \label{eq:S-qutrit-beta-majorant}
\end{equation}
Write the nodes as $c_j=2j+\epsilon$, where $\epsilon$ is the parity of $s$, and let $e_j=c_j+1$. Equation~\eqref{eq:S-jacobi-gamma} then gives
\begin{equation}
 \gamma_{e_j}=(b_j^{(\epsilon)})^2q_{c_j}\le(b_j^{(\epsilon)})^2q_*.
 \label{eq:S-qutrit-gamma-model}
\end{equation}
Thus the ordinary off-diagonal part of the Jacobi matrix is entrywise dominated by $\frac12B(q_*)\sqrt{q_*}\,\mathsf K_{\epsilon,s}$. When $s$ is odd, the one-sided coefficient has $\rho_0=(r-1)/2$ and shell parameter
\begin{equation}
 \widetilde q_0=\frac{r-1}{8(2n-4t+r+1)}\le q_*,
\end{equation}
so $\beta_0\le B(q_*)$. The corresponding one-sided weight is
\begin{equation}
 B_0=\frac{2D_1}{7(s+1)}.
 \label{eq:S-qutrit-B0}
\end{equation}
Hence its diagonal load is at most $2B(q_*)/[7(s+1)]$. Using the parity-model norm $3/\sqrt2$ from Sec.~\ref{sec:Sjacobi} and bounding this one-sided diagonal term separately yields
\begin{equation}
 \frac{C_s}{I_s}\le A(q_*)+B(q_*)\left(\frac{3}{2\sqrt2}\sqrt{q_*}+\varepsilon_s\frac{2}{7(s+1)}\right),
 \label{eq:S-qutrit-model-bound}
\end{equation}
where $\varepsilon_s=1$ for odd $s$ and $\varepsilon_s=0$ for even $s$.

We now retain the integer parameters in the scalar comparison. The long-range hypothesis gives
\begin{equation}
 q_*\le q_{t,r}:=\frac{r}{8\left((205/42)t+r+2\right)}<\frac18.
 \label{eq:S-qutrit-qtr}
\end{equation}
Also,
\begin{equation}
 \frac{D_n(t,s)}{t}\ge d_{t,r}:=\frac{746}{21}-\frac{9(2t-r-1)}{t\left(1-\kappa(r/t)\right)}.
 \label{eq:S-qutrit-dtr}
\end{equation}
Since $2t-r-1=s-1<2t$ and $\kappa(r/t)\le1/h$,
\begin{equation}
 d_{t,r}>\frac{746}{21}-\frac{18}{1-1/h}=\frac{368}{21}-3\sqrt2>9.
 \label{eq:S-qutrit-dtr-positive}
\end{equation}
The map $d\mapsto d^2/[9(2d-9)]$ is increasing for $d>9$, so
\begin{equation}
 T_n(t,s)\ge\Theta_{t,r}:=\frac{d_{t,r}^2}{9(2d_{t,r}-9)}.
 \label{eq:S-qutrit-Thetatr}
\end{equation}
The functions $A(q)$, $B(q)$, and $\sqrt q$ are increasing on $0\le q<1$. Hence it is enough to prove
\begin{align}
 \Theta_{t,r}&\ge A(q_{t,r})+B(q_{t,r})\frac{3}{2\sqrt2}\sqrt{q_{t,r}}, && s\ {\rm even},
 \label{eq:S-qutrit-scalar-even}\\
 \Theta_{t,r}&\ge A(q_{t,r})+B(q_{t,r})\left(\frac{3}{2\sqrt2}\sqrt{q_{t,r}}+\frac{2}{7(s+1)}\right), && s\ {\rm odd}.
 \label{eq:S-qutrit-scalar-odd}
\end{align}
The parity reduction and exact symbolic verification of Eqs.~\eqref{eq:S-qutrit-scalar-even} and \eqref{eq:S-qutrit-scalar-odd} are given together with the code in the Appendix. Combining these inequalities with Eqs.~\eqref{eq:S-qutrit-target} and \eqref{eq:S-qutrit-model-bound} proves, for $t\ge2$,
\begin{equation}
 R_{9;n,t}(2)\ge R_{9;n,t}(3)\ge\cdots\ge R_{9;n,t}(2t).
 \label{eq:S-qutrit-chain}
\end{equation}

It remains to control the endpoints. Put
\begin{equation}
 \vartheta=\frac{t}{8(n-t-1)}\le\frac{21}{578},
 \label{eq:S-qutrit-vartheta}
\end{equation}
where the bound follows from $n-t-1\ge(289/84)t$. For separation one, let $\vartheta_1=t/[8(n-t)]<\vartheta$ and $W_1=w_t^{(9)}(n-1)$. Equation~\eqref{eq:S-shell-tail} gives
\begin{equation}
 \Vol_{t-1}(n-1)\le\frac{\vartheta_1W_1}{1-\vartheta_1},\qquad \Vol_t(n-1)\le\frac{W_1}{1-\vartheta_1}.
\end{equation}
Since $c_t^{(9)}(n,1)=9\Vol_{t-1}(n-1)$, $\Vol_t(n)=\Vol_t(n-1)+8\Vol_{t-1}(n-1)$, and $\Lloyd_t^{(9,n)}(1)=W_1$,
\begin{equation}
 R_{9;n,t}(1)\le\frac{9\vartheta_1(1+8\vartheta_1)}{(1-\vartheta_1)^2}.
 \label{eq:S-qutrit-R1-theta}
\end{equation}
For separation two and $t\ge2$, put $M=n-2$ and $W_2=w_t^{(9)}(M)$. The shell-tail estimate gives
\begin{equation}
 \Vol_t(M)\le\frac{W_2}{1-\vartheta},\qquad \Vol_{t-1}(M)\le\frac{\vartheta W_2}{1-\vartheta},\qquad \Vol_{t-2}(M)\le\frac{\vartheta^2W_2}{1-\vartheta}.
 \label{eq:S-qutrit-endpoint-shells}
\end{equation}
Using $\Vol_t(n)=\Vol_t(M)+16\Vol_{t-1}(M)+64\Vol_{t-2}(M)$ and the direct count
\begin{equation}
 c_t^{(9)}(n,2)=2\Vol_{t-1}(M)+79\Vol_{t-2}(M),
\end{equation}
we obtain
\begin{equation}
 \Vol_t(n)\le\frac{(1+8\vartheta)^2W_2}{1-\vartheta},\qquad c_t^{(9)}(n,2)\le\frac{(2\vartheta+79\vartheta^2)W_2}{1-\vartheta}.
\end{equation}
Moreover,
\begin{equation}
 \Lloyd_t^{(9,n)}(2)=P_t^{(9)}(1;n-1)=W_2-w_{t-1}^{(9)}(M)=(1-\vartheta)W_2.
\end{equation}
Therefore
\begin{equation}
 R_{9;n,t}(2)\le\frac{(1+8\vartheta)^2(2\vartheta+79\vartheta^2)}{(1-\vartheta)^4}\qquad(t\ge2).
 \label{eq:S-qutrit-R2-theta}
\end{equation}
The two right-hand sides in Eqs.~\eqref{eq:S-qutrit-R1-theta} and \eqref{eq:S-qutrit-R2-theta} are increasing for $0\le\vartheta<1$, since
\begin{align}
 \frac{d}{d\vartheta}\frac{9\vartheta(1+8\vartheta)}{(1-\vartheta)^2}&=\frac{9(17\vartheta+1)}{(1-\vartheta)^3}>0,\\
 \frac{d}{d\vartheta}\frac{(1+8\vartheta)^2(2\vartheta+79\vartheta^2)}{(1-\vartheta)^4}&=\frac{2(8\vartheta+1)(1351\vartheta^2+106\vartheta+1)}{(1-\vartheta)^5}>0.
\end{align}
Since $\vartheta_1<\vartheta\le21/578$, these monotonicity formulas give
\begin{equation}
 R_{9;n,t}(1)\le\frac{140994}{310249}<1,
 \label{eq:S-qutrit-R1}
\end{equation}
and, for $t\ge2$,
\begin{equation}
 R_{9;n,t}(2)\le\frac{32898443340}{96254442001}<1.
 \label{eq:S-qutrit-R2}
\end{equation}
When $t=1$, the long-range condition forces $n\ge6$. Here $c_1^{(9)}(n,2)=2$, $\Vol_1(n)=8n+1$, and $\Lloyd_1^{(9,n)}(2)=8n-17$, so
\begin{equation}
 R_{9;n,1}(2)=\frac{2(8n+1)}{(8n-17)^2}<1,
\end{equation}
because $(8n-17)^2-2(8n+1)=32n(2n-9)+287>0$ for $n\ge6$.

For $t\ge2$, Eqs.~\eqref{eq:S-qutrit-chain}, \eqref{eq:S-qutrit-R1}, and \eqref{eq:S-qutrit-R2} give $R_{9;n,t}(s)<1$ for every $1\le s\le2t$; for $t=1$, Eqs.~\eqref{eq:S-qutrit-R1} and the direct separation-two calculation give the same conclusion. Together with the Lloyd positivity proved above, the raw Lloyd-square defects satisfy
\begin{equation}
 \Psi_{t,n,s}:=\Lloyd_t^{(9,n)}(s)^2-\Vol_t(n)c_t^{(9)}(n,s)\ge0
 \label{eq:S-qutrit-long-psi}
\end{equation}
for every $1\le s\le2t$ and every $n-1\ge(373/84)t$.

\smsection{Filtered qutrit bridge}{sec:S5}

Throughout this section assume
\begin{equation}
 \frac{72}{17}t<n<1+\frac{373}{84}t,\qquad t\ge2.
 \label{eq:S-bridge-repeat}
\end{equation}
We use the quadratic multiplier
\begin{equation}
 h_{n,t}(i)=(n-i)(9i+n+7t-1),\qquad f_i=h_{n,t}(i)\Lloyd_t^{(9,n)}(i)^2.
 \label{eq:S-filter}
\end{equation}
Since $0\le i\le n$ and $9i+n+7t-1>0$, one has $h_{n,t}(i)\ge0$, so the Fourier coefficients $f_i$ are nonnegative.

Let $w_t=8^t\binom nt$. The probability measure on Fourier shells induced by the qutrit Lloyd square is
\begin{equation}
 \mu_t(i)=\frac{8^i\binom ni\Lloyd_t^{(9,n)}(i)^2}{9^n\Vol_t(n)},\qquad0\le i\le n,
 \label{eq:S-mu-def}
\end{equation}
where the denominator is Parseval's identity for $1_{B_t}$. Equivalently, if
\begin{equation}
 M(r)=\sum_{i=0}^n8^i\binom ni\Lloyd_t^{(9,n)}(i)^2r^i,
\end{equation}
then $M(1)=9^n\Vol_t(n)$. The first two shell-marker derivatives are
\begin{align}
 M'(1)&=9^{n-1}8(n-t)w_t,
 \label{eq:S-Mprime}\\
 M''(1)+M'(1)&=9^{n-2}8(n-t)(8n-7t+1)w_t.
 \label{eq:S-Msecond}
\end{align}
They follow by differentiating the ball autocorrelation generating function with one, respectively two, marked Fourier coordinates; equivalently, by Krawtchouk orthogonality after inserting the shell factors $i$ and $i(i-1)$. Hence
\begin{equation}
 \mathbb E_{\mu_t}i=\frac{8(n-t)w_t}{9\Vol_t(n)},\qquad \mathbb E_{\mu_t}i^2=\frac{8(n-t)(8n-7t+1)w_t}{81\Vol_t(n)},
 \label{eq:S-filter-moments}
\end{equation}
and therefore
\begin{equation}
 \mathbb E_{\mu_t}\{(8n-7t+1)i-9i^2\}=0.
 \label{eq:S-moment-cancel}
\end{equation}
Since
\begin{equation}
 h_{n,t}(i)-h_{n,t}(0)=(8n-7t+1)i-9i^2,
\end{equation}
we have $\mathbb E_{\mu_t}h_{n,t}=h_{n,t}(0)$. Because $\Lloyd_t^{(9,n)}(0)=\Vol_t(n)$, the zeroth Krawtchouk coefficient of the filtered square is $f_0=h_{n,t}(0)\Vol_t(n)^2$, whereas its distance-space value at the origin is $F(0)=9^n\Vol_t(n)h_{n,t}(0)$. Thus
\begin{equation}
 \frac{F(0)}{f_0}=\frac{9^n}{\Vol_t(n)}.
 \label{eq:S-prefactor}
\end{equation}
The filter therefore preserves the Hamming prefactor required by the Li--Xing comparison.

We now pass to distance shells. Write
\begin{equation}
 C_s=c_t^{(9)}(n,s),\qquad L_s=\Lloyd_t^{(9,n)}(s),\qquad H_s=h_{n,t}(s),
 \label{eq:S-bridge-CLH}
\end{equation}
and define
\begin{equation}
 \Gamma_s=9^{-n}\sum_{i=0}^{n}H_iL_i^2\Kraw_i^{(9)}(s;n).
 \label{eq:S-filter-gamma}
\end{equation}
With the normalization above, the Li--Xing defect on an overlapping shell is
\begin{equation}
 \Phi_{n,t,s}=H_sL_s^2-\Vol_t(n)\Gamma_s,
 \label{eq:S-filter-defect-def}
\end{equation}
so the required overlap condition is
\begin{equation}
 \Phi_{n,t,s}\ge0,\qquad1\le s\le2t.
 \label{eq:S-filter-inside-target}
\end{equation}

Let $\mathcal T$ be the unnormalized Hamming adjacency operator on radial functions,
\begin{equation}
 (\mathcal TF)_s=sF_{s-1}+7sF_s+8(n-s)F_{s+1},
 \label{eq:S-hamming-adjacency}
\end{equation}
with out-of-range terms omitted. Since $\Kraw_1^{(9)}(i;n)=8n-9i$, multiplication by $8n-9i$ on the Fourier side is $\mathcal T$ on distance shells. Expanding $H_i=(n-i)(9i+n+7t-1)$ gives
\begin{equation}
 \Gamma_s=\frac19\left\{n(9n+7t-1)C_s+(8n+7t-1)(\mathcal TC)_s-(\mathcal T^2C)_s\right\}.
 \label{eq:S-filter-gamma-adjacency}
\end{equation}
The second iterate is
\begin{align}
 (\mathcal T^2C)_s={}&s(s-1)C_{s-2}+7s(2s-1)C_{s-1}+\{8n(2s+1)+33s^2\}C_s\nonumber\\
 &+56(n-s)(2s+1)C_{s+1}+64(n-s)(n-s-1)C_{s+2}.
 \label{eq:S-filter-T2}
\end{align}
Since $C_j=0$ for $j>2t$, these identities immediately give
\begin{equation}
 \Gamma_s=0\quad(s\ge2t+3),\qquad \Gamma_{2t+2}=-\frac{(2t+2)(2t+1)}9C_{2t}<0.
 \label{eq:S-filter-tail-2}
\end{equation}
Moreover,
\begin{equation}
 C_{2t}=\binom{2t}{t},\qquad C_{2t-1}=\left(1+\frac{7t}{2}\right)\binom{2t}{t},
 \label{eq:S-endpoint-intersections}
\end{equation}
so
\begin{equation}
 \Gamma_{2t+1}=\frac{(2t+1)C_{2t}}9\bigl(8n-7t^2-23t-8\bigr).
 \label{eq:S-filter-tail-1b}
\end{equation}
In the bridge and for $t\ge2$,
\begin{equation}
 8n<8+\frac{746}{21}t\le7t^2+23t+8,
 \label{eq:S-filter-tail-bridge-bound}
\end{equation}
whence $\Gamma_{2t+1}<0$. Therefore $9^n\Gamma_s\le0$ on every nonoverlapping shell $s>2t$.

For overlapping shells, the two-center intersection has the coefficient form
\begin{equation}
 C_s=[u^{\le t}v^{\le t}](u+v+7uv)^s(1+8uv)^{n-s}.
 \label{eq:S-two-var-intersection}
\end{equation}
Put
\begin{equation}
 I_s=C_s-C_{s+1},\qquad0\le s\le2t.
 \label{eq:S-increment-def}
\end{equation}
The difference retains the balance between the two centers.

\begin{lemma}[Balanced increment]
\label{lem:S-balanced-increment}
For $0\le s\le2t$,
\begin{equation}
 I_s=[u^tv^t](u+v+7uv)^s(1+8uv)^{n-s-1}.
 \label{eq:S-balanced-increment}
\end{equation}
Equivalently,
\begin{equation}
 I_s=\sum_{k=\max(0,s-t)}^{\lfloor s/2\rfloor}\binom{s}{k}\binom{s-k}{k}7^{s-2k}8^{t-s+k}\binom{n-s-1}{t-s+k}.
 \label{eq:S-balanced-increment-sum}
\end{equation}
In particular $I_s\ge0$ and, since $C_{2t+1}=0$,
\begin{equation}
 C_s=\sum_{j=s}^{2t}I_j.
 \label{eq:S-C-suffix}
\end{equation}
\end{lemma}

\begin{proof}
Set $a=u+v+7uv$ and $b=1+8uv$. Subtracting the coefficient forms for $C_s$ and $C_{s+1}$ gives
\begin{equation}
 I_s=[u^{\le t}v^{\le t}]a^sb^{n-s-1}(b-a).
\end{equation}
Since $b-a=(1-u)(1-v)$, the two-variable telescoping identity
\begin{equation}
 [u^{\le t}v^{\le t}](1-u)(1-v)P=[u^tv^t]P
\end{equation}
for a polynomial $P$ with nonnegative powers proves Eq.~\eqref{eq:S-balanced-increment}. For Eq.~\eqref{eq:S-balanced-increment-sum}, choose $k$ $u$-only factors and $k$ $v$-only factors from $a^s$; the remaining $s-2k$ factors contribute $7uv$, and $b^{n-s-1}$ supplies $t-s+k$ further copies of $uv$.
\end{proof}

The filtered adjacency now collapses to the same increments. Substituting
\begin{equation}
 C_{s-2}=C_s+I_{s-2}+I_{s-1},\qquad C_{s-1}=C_s+I_{s-1},\qquad C_{s+1}=C_s-I_s,\qquad C_{s+2}=C_s-I_s-I_{s+1}
\end{equation}
into Eqs.~\eqref{eq:S-filter-gamma-adjacency} and \eqref{eq:S-filter-T2} gives
\begin{align}
 \Gamma_s={}&n(n+7t-1)C_s-\frac{s(s-1)}9I_{s-2}\notag\\
 &+\frac{s(8n-15s+7t+7)}9I_{s-1}+\frac{8(n-s)(6s-7t)}9I_s+\frac{64(n-s)(n-s-1)}9I_{s+1}.
 \label{eq:S-gamma-increments}
\end{align}
For $s=1$ the coefficient of $I_{s-2}$ vanishes, so no negative-index value is needed. Together with Eq.~\eqref{eq:S-balanced-increment}, Eq.~\eqref{eq:S-gamma-increments} is the exact bridge identity used below.

To place all terms on one parity chain, set
\begin{equation}
 r=2t-s,\qquad N=n-s-1.
 \label{eq:S-bridge-rN}
\end{equation}
Let
\begin{equation}
 \mathcal C_s=\{c:0\le c\le\min(s,r),\ c\equiv s\pmod2\},\qquad m_c=s-c,\qquad \ell_c=\frac{r-c}{2}.
\end{equation}
For $c\in\mathcal C_s$, put
\begin{equation}
 D_c=\binom{s}{c}7^c\binom{m_c}{m_c/2}8^{\ell_c}\binom{N}{\ell_c}.
 \label{eq:S-bridge-Dc}
\end{equation}
Then Eq.~\eqref{eq:S-balanced-increment-sum} becomes
\begin{equation}
 I_s=\sum_{c\in\mathcal C_s}D_c.
 \label{eq:S-bridge-I-nodes}
\end{equation}
For the opposite parity define
\begin{equation}
 \mathcal E_s=\{e:1\le e\le\min(s,r-1),\ e\not\equiv s\pmod2\},\qquad m_e=s-e,\qquad \rho_e=\frac{r-e-1}{2},
\end{equation}
and, for $e\in\mathcal E_s$,
\begin{equation}
 B_e=\binom{s}{e}7^e\binom{m_e}{(m_e-1)/2}8^{\rho_e}\binom{N}{\rho_e}.
 \label{eq:S-bridge-Be}
\end{equation}
These are the $Q=9$ specializations of the node and ordinary-edge weights in Sec.~\ref{sec:Sjacobi}. With the corresponding coefficients $\alpha_c$ and $\beta_e$ from Eqs.~\eqref{eq:S-jacobi-alpha} and \eqref{eq:S-jacobi-beta}, the exact two-ball decomposition is
\begin{equation}
 C_s=\sum_{c\in\mathcal C_s}\alpha_cD_c+\sum_{e\in\mathcal E_s}\beta_eB_e+\mathbf1_{\{s\ {\rm odd}\}}\beta_0B_0,
 \label{eq:S-bridge-node-edge-C}
\end{equation}
where, when $s$ is odd,
\begin{equation}
 \rho_0=\frac{r-1}{2},\qquad B_0=\binom{s}{(s-1)/2}8^{\rho_0}\binom{N}{\rho_0}=\frac{2D_1}{7(s+1)}.
 \label{eq:S-bridge-B0}
\end{equation}
Thus the one-sided term is kept explicitly rather than absorbed into a node coefficient.

The adjacent increments in Eq.~\eqref{eq:S-gamma-increments} lie on the same chain. On common support, direct binomial cancellation gives
\begin{align}
 R_c^{(-2)}&=\frac{2m_c^2(N+1)(N+2)}{s(s-1)(\ell_c+1)(N-\ell_c+1)},
 \label{eq:S-transfer-m2}\\
 R_e^{(-1)}&=\frac{4(m_e+1)(N+1)}{s(\rho_e+1)},
 \label{eq:S-transfer-m1}\\
 R_e^{(+1)}&=\frac{2(s+1)(N-\rho_e)}{N(m_e+1)}.
 \label{eq:S-transfer-p1}
\end{align}
Accordingly, the common-support parts of $I_{s-2}$, $I_{s-1}$, and $I_{s+1}$ are obtained by multiplying $D_c$, $B_e$, and $B_e$, respectively, by these ratios. Only endpoint mismatches remain. If $r\le s-2$, the additional right-end term $E_{-1}$ of $I_{s-1}$ satisfies
\begin{equation}
 \frac{E_{-1}}{D_r}=\frac{7(s-r)^2}{4s(r+1)}.
 \label{eq:S-transfer-end-m1}
\end{equation}
If $r\ge s+2$, the additional right-end term $E_{+1}$ of $I_{s+1}$ satisfies
\begin{equation}
 \frac{E_{+1}}{D_s}=\frac{7\ell_s}{8N}.
 \label{eq:S-transfer-end-p1}
\end{equation}
The unmatched endpoint of $I_{s-2}$ is nonnegative and carries the negative coefficient $-s(s-1)/9$ in Eq.~\eqref{eq:S-gamma-increments}, so it may be retained as favorable slack or omitted in an upper bound. These exact transfer identities place the filtered defect on the same balance-preserving parity chain as the two-center intersection.

We now use the exact identities above to prove Eq.~\eqref{eq:S-filter-inside-target}. Put
\begin{equation}
 \text{core: }200r\ge29t,\qquad \text{tail: }200r<29t,\qquad r=2t-s.
 \label{eq:S-core-tail-split}
\end{equation}
The core retains enough raw Lloyd-square margin to absorb the quadratic filter. In the tail, the filtered distance-space value is nonpositive by a direct increment-ratio estimate.

\begin{lemma}[Core half-gap]
\label{lem:S-core-half-gap}
For every bridge triple with $t\ge10$ and every overlapping shell $1\le s\le2t$ satisfying $200(2t-s)\ge29t$,
\begin{equation}
 L_s^2>2\Vol_t(n)C_s.
 \label{eq:S-core-half-gap}
\end{equation}
\end{lemma}

\begin{proof}
For an ordinary edge, direct cancellation in Eqs.~\eqref{eq:S-bridge-Dc} and \eqref{eq:S-bridge-Be} gives
\begin{equation}
 \frac{B_e}{D_{e+1}}=\frac{2(e+1)}{7(m_e+1)},\qquad \frac{B_e}{D_{e-1}}=\frac{7(m_e+1)(\rho_e+1)}{16e(N-\rho_e)}.
 \label{eq:S-edge-two-ratios}
\end{equation}
Hence
\begin{equation}
 z_e:=\frac{B_e}{D_{e-1}+D_{e+1}}=\frac{14(e+1)(m_e+1)(\rho_e+1)}{32e(e+1)(N-\rho_e)+49(m_e+1)^2(\rho_e+1)}.
 \label{eq:S-proportional-charge}
\end{equation}
For the one-sided odd edge, $z_0=2/[7(s+1)]$. Charging each $B_e$ proportionally to its neighboring nodes makes Eq.~\eqref{eq:S-bridge-node-edge-C} an exact node sum. With
\begin{equation}
 q_* =\frac{r}{8(2n-4t+r)},\qquad A(q)=\frac{(1+8q)(1+q)}{(1-q)^2},\qquad B(q)=\frac{2(1+8q)}{(1-q)^2},
 \label{eq:S-core-AB}
\end{equation}
the shell-tail bounds from Sec.~\ref{sec:S4} give
\begin{equation}
 \frac{C_s}{I_s}\le A(q_*)+B(q_*)\max_c\{z_{c-1}+z_{c+1}\}.
 \label{eq:S-core-charge-bound}
\end{equation}
Absent edges contribute zero, with $z_0$ used at the odd left endpoint.

For $2\le s\le2t-1$, set
\begin{equation}
 \kappa=\frac{2(t-1)}{8n-25t+9r+8}.
 \label{eq:S-core-kappa}
\end{equation}
The degree recurrence for $P_j=\Kraw_j^{(9)}(s-2;n-3)$ gives $0<P_{j-1}/P_j\le\kappa$ for $1\le j\le t-1$ once
\begin{equation}
 (8n-25t+9r+8)^2\ge32(t-1)(n-t).
 \label{eq:S-core-discriminant}
\end{equation}
Indeed, the induction residual decreases with $j$, and its value at $j=t-1$ is
\begin{equation}
 (t-1)\left(1-\frac{32(t-1)(n-t)}{(8n-25t+9r+8)^2}\right).
\end{equation}
On $200r\ge29t$, the left factor in the square is at least $(34637/3400)t+8$ at $n=72t/17$, and the leading margin in Eq.~\eqref{eq:S-core-discriminant} is $2921769t^2/11560000>0$. The margin increases with $n$.

For $t\ge10$, $\kappa\le6800/34637$ and $D_n(t,s)/t\ge129445081/10516200>9$. With $\rho=P_{t-2}/P_{t-1}$, the preceding ratio bound gives $0<\rho\le\kappa<1$. The shifts \eqref{eq:S-qutrit-shift-M}--\eqref{eq:S-qutrit-shift-L} therefore give $M_s>0$ and $L_s/M_s\ge D_n(t,s)/t>9$. Since $L_{s+1}=L_s-9M_s$, it follows that $L_s,L_{s+1}>0$.

The Lloyd contiguous step from Sec.~\ref{sec:S4} now gives
\begin{equation}
 \frac{C_s}{I_s}\le T_n(t,s)\quad\Longrightarrow\quad R_{9;n,t}(s+1)\le R_{9;n,t}(s),
 \label{eq:S-core-monotone-criterion}
\end{equation}
where
\begin{equation}
 D_n(t,s)=8(n-1)-\frac{9(s-1)}{1-\kappa},\qquad T_n(t,s)=\frac{D_n(t,s)^2}{9t(2D_n(t,s)-9t)}.
 \label{eq:S-core-T}
\end{equation}

It remains to compare Eq.~\eqref{eq:S-core-charge-bound} with Eq.~\eqref{eq:S-core-T}. The functions $A$ and $B$ are increasing on $0\le q<1$. As $n$ increases, $q_*$ and every ordinary $z_e$ decrease, while the one-sided value $z_0=2/[7(s+1)]$ is independent of $n$. Thus the right side of Eq.~\eqref{eq:S-core-charge-bound} is nonincreasing in $n$. At the same time $D_n(t,s)$ increases, and $T_n$ is increasing in $D_n$ once $D_n>9t$. Therefore the comparison is weakest at the real endpoint $n=72t/17$. The remaining rational inequality on the physical core is certified exactly by the Bernstein calculation recorded in the Appendix.

The two starting shells have direct bounds. From $n>72t/17$, $\vartheta=t/[8(n-t)]<17/440$, so
\begin{equation}
 R_{9;n,t}(1)\le\frac{9\vartheta(1+8\vartheta)}{(1-\vartheta)^2}<\frac{1088}{2209}<\frac12.
 \label{eq:S-core-R1}
\end{equation}
For $t\ge10$, with $\eta=t/[8(n-t-1)]\le85/2132$,
\begin{equation}
 R_{9;n,t}(2)\le\frac{(1+8\eta)^2(2\eta+79\eta^2)}{(1-\eta)^4}\le\frac{7379252030960}{17557851463681}<\frac12.
 \label{eq:S-core-R2}
\end{equation}
The monotonicity implication \eqref{eq:S-core-monotone-criterion} propagates the $s=2$ bound through every later core shell, while Eq.~\eqref{eq:S-core-R1} handles $s=1$. This proves Eq.~\eqref{eq:S-core-half-gap}.
\end{proof}

\begin{lemma}[Core filtered multiplier]
\label{lem:S-core-multiplier}
For every overlapping bridge shell $1\le s\le2t$ with $200(2t-s)\ge29t$ and $t\ge100$,
\begin{equation}
 \Gamma_s\le2H_sC_s.
 \label{eq:S-core-multiplier}
\end{equation}
\end{lemma}

\begin{proof}
Write $h_0=n(n+7t-1)$. The coefficient $h_0-2H_s$ is negative throughout the bridge. Indeed it is decreasing in $n$ and convex in $s$, while at $n=4t+1$ its endpoint values are $-t(44t+61)$ for $s=1$ and $-t(72t+47)$ for $s=2t$. Since $C_s\ge I_s=\sum_cD_c$, we may therefore replace $(h_0-2H_s)C_s$ by the larger quantity $(h_0-2H_s)I_s$.

Insert the transfer ratios \eqref{eq:S-transfer-m2}--\eqref{eq:S-transfer-p1} into the remaining terms of Eq.~\eqref{eq:S-gamma-increments} and distribute each $B_e$ proportionally between its adjacent $D$-nodes as in Eq.~\eqref{eq:S-proportional-charge}. The unmatched $I_{s-2}$ endpoint is favorable because its coefficient is negative, and the endpoint terms \eqref{eq:S-transfer-end-m1}--\eqref{eq:S-transfer-end-p1} agree with the same edge formula under the one-sided endpoint convention. Hence it is enough to show a local rational residual at every node.

Put
\begin{equation}
 x=\frac1t,\qquad u=\frac rt,\qquad \sigma=\frac st=2-u,\qquad w=\frac ct,\qquad \lambda=\frac nt.
 \label{eq:S-mult-vars}
\end{equation}
For a neighboring edge coordinate $v=e/t$, set
\begin{align}
 \nu(v)&=\lambda-2+\frac{u+v-x}{2},\nonumber\\
 \mathcal D(v)&=32v(v+x)\nu(v)+\frac{49}{2}(2-u-v+x)^2(u-v+x),
 \label{eq:S-mult-D}\\
 P_-(v)&=\frac{56(v+x)(2-u-v+x)^2(\lambda-2+u)}{(2-u)\mathcal D(v)},\nonumber\\
 P_+(v)&=\frac{14(v+x)(u-v+x)(2-u+x)\nu(v)}{(\lambda-2+u-x)\mathcal D(v)}.
 \label{eq:S-mult-P}
\end{align}
All displayed denominators are positive at a physical bridge node; the same expressions include the one-sided and additional endpoint terms. Normalize the coefficients in Eq.~\eqref{eq:S-gamma-increments} by $t^2$ and write
\begin{align}
 \delta={}&\lambda(\lambda+7-x)-2(\lambda-\sigma)(9\sigma+\lambda+7-x),\nonumber\\
 A_0={}&\frac89(\lambda-\sigma)(6\sigma-7),\nonumber\\
 A_{-1}={}&\frac{\sigma(8\lambda-15\sigma+7+7x)}9,\nonumber\\
 A_{+1}={}&\frac{64}{9}(\lambda-\sigma)(\lambda-\sigma-x),\nonumber\\
 E_{-2}={}&-\frac{2(\sigma-w)^2(\lambda-2+u)(\lambda-2+u+x)}{9((u-w)/2+x)(\lambda-2+(u+w)/2)}.
 \label{eq:S-mult-coeff}
\end{align}
The coefficient of a node in $t^{-2}(\Gamma_s-2H_sC_s)$ is bounded above by
\begin{equation}
 U=\delta+A_0+E_{-2}+A_{-1}\{P_-(w-x)+P_-(w+x)\}+A_{+1}\{P_+(w-x)+P_+(w+x)\}.
 \label{eq:S-mult-U}
\end{equation}
Thus $U\le0$ at every physical node proves Eq.~\eqref{eq:S-core-multiplier}. The endpoint $s=1$ is the corresponding one-node specialization. The exact Bernstein certificates for the late core, early core, $c=0$ faces, and $s=1$ are recorded together with the code in the Appendix, and they give $U\le0$ throughout the physical core.
\end{proof}

Combining Lemmas~\ref{lem:S-core-half-gap} and \ref{lem:S-core-multiplier}, for every analytic core shell $t\ge100$,
\begin{equation}
 \Vol_t(n)\Gamma_s\le2H_s\Vol_t(n)C_s<H_sL_s^2,
 \label{eq:S-core-defect-positive}
\end{equation}
so $\Phi_{n,t,s}>0$.

\begin{lemma}[Tail increment ratio]
\label{lem:S-tail-ratio}
Let $t\le j\le2t$, set $I_{2t+1}:=0$, and put $R=2t-j$. Then
\begin{equation}
 \frac{I_{j+1}}{I_j}\le Q_t(R):=\frac{R(2t-R+1)}{7(t-R+1)^2}.
 \label{eq:S-tail-ratio}
\end{equation}
\end{lemma}

\begin{proof}
For $j=2t$ the assertion is immediate from $I_{2t+1}=0$. For $t\le j\le2t-1$, use Eq.~\eqref{eq:S-balanced-increment-sum} and pair a term of $I_{j+1}$ indexed by $k$ with the term of $I_j$ indexed by $k-1$. Before the outside-binomial factor is bounded, the ratio is
\begin{equation}
 \frac{(j+1)(j-2k+2)(N-\ell)}{7k^2N},
\end{equation}
with the corresponding outside-shell index $\ell$. The support constraints give $j-2k+2\le R$, $k\ge t-R+1$, and $(N-\ell)/N\le1$, while $j+1=2t-R+1$. This yields Eq.~\eqref{eq:S-tail-ratio} term by term and hence after summation.
\end{proof}

Set
\begin{equation}
 q=\frac9{160}.
 \label{eq:S-tail-q}
\end{equation}
On $200r<29t$ and $t\ge225$, the three indices $j=s-2,s-1,s$ all satisfy $j\ge t$ and have $R\le r+2$. The function $Q_t(R)$ is increasing in the relevant range, since
\begin{equation}
 Q_t'(R)=\frac{2t^2+3t+1-R}{7(t-R+1)^3}>0.
\end{equation}
At the relaxed endpoint $R=29t/200+2$,
\begin{equation}
 7q(t-R+1)^2-R(2t-R+1)=\frac{120743t^2-27125200t+15320000}{6400000}>0
 \label{eq:S-tail-q-margin}
\end{equation}
for $t\ge225$. Hence all three adjacent ratios are $<q$.

Write the coefficients in Eq.~\eqref{eq:S-gamma-increments} as $h_0=n(n+7t-1)$, $-b=-s(s-1)/9$, and $a_{-1},a_0,a_1$ in their displayed order. In the tail,
\begin{equation}
 a_{-1}-\frac bq=-\frac{s(-72n+295s-63t-223)}{81}\le0,
 \label{eq:S-tail-pair-sign}
\end{equation}
because at the bridge upper length and tail lower separation
\begin{equation}
 -72n+295s-63t-223>\frac{46063t-82600}{280}>0.
\end{equation}
For every $s\le j\le2t-1$, one has $2t-j\le r$, and hence $I_{j+1}/I_j\le Q_t(2t-j)\le Q_t(r)<q$. The entire suffix is therefore controlled by the same geometric ratio:
\begin{equation}
 C_s=\sum_{j=s}^{2t}I_j\le\frac{I_s}{1-q}.
\end{equation}
The adjacent bounds at $j=s-2,s-1,s$, together with Eq.~\eqref{eq:S-tail-pair-sign}, also give
\begin{equation}
 -bI_{s-2}+a_{-1}I_{s-1}\le\left(\frac{a_{-1}}q-\frac b{q^2}\right)I_s,\qquad I_{s+1}\le qI_s.
\end{equation}
Therefore
\begin{equation}
 \frac{\Gamma_s}{I_s}\le\mathcal G:=\frac{h_0}{1-q}+a_0+a_1q+\frac{a_{-1}}q-\frac b{q^2}.
 \label{eq:S-tail-G}
\end{equation}
Exact simplification gives
\begin{align}
 \mathcal G=-\frac{58}{550395}\bigl(-13851n^2-192978ns-11340nt+13851n+661229s^2-190260st-468251s\bigr).
 \label{eq:S-tail-G-poly}
\end{align}
After putting $s=2t-r$, this upper bound is increasing in both $n$ and $r$ on the tail. Indeed,
\begin{align}
 \partial_n\mathcal G&=-\frac{58}{61155}(-3078n+21442r-44144t+1539)>0,\nonumber\\
 \partial_r\mathcal G&=-\frac{58}{3645}(1278n+8758r-16256t+3101)>0.
 \label{eq:S-tail-monotonicity}
\end{align}
The two parenthesized factors are negative: using $n>4t$ and $r<29t/200$ for the first, and the two bridge upper bounds for the second,
\begin{align}
 -3078n+21442r-44144t+1539&<-\frac{5334691}{100}t+1539<0,\notag\\
 1278n+8758r-16256t+3101&<-\frac{6517813}{700}t+4379<0
\end{align}
for $t\ge225$. Consequently the largest value is at the relaxed boundary $n=1+373t/84$, $r=29t/200$, where
\begin{equation}
 \mathcal G=-\frac{29t(18288182861t-2546872668200)}{539387100000}<0\qquad(t\ge225).
 \label{eq:S-tail-G-boundary}
\end{equation}
Thus
\begin{equation}
 \Gamma_s\le0\qquad(t\ge225,\;200(2t-s)<29t),
 \label{eq:S-tail-Gamma-negative}
\end{equation}
and the filtered defect is automatically nonnegative there.

The remaining finite set consists of every bridge shell with $2\le t<100$ and the tail shells with $100\le t<225$. The exact finite certificate in the Appendix evaluates all $251602$ triples and gives
\begin{equation}
 \min\Phi_{n,t,s}=\Phi_{9,2,4}=\frac{224582966}{3}>0.
 \label{eq:S-finite-min-Phi}
\end{equation}

\begin{theorem}[Filtered qutrit bridge]
\label{thm:S-qutrit-bridge}
For every integer triple in the bridge window \eqref{eq:S-bridge} with $t\ge2$, the filtered Lloyd square \eqref{eq:S-filter} satisfies
\begin{equation}
 \Phi_{n,t,s}\ge0\quad(1\le s\le2t),\qquad \Gamma_s\le0\quad(s>2t).
 \label{eq:S-filtered-bridge-final}
\end{equation}
\end{theorem}

\begin{proof}
For $s>2t$ this is Eqs.~\eqref{eq:S-filter-tail-2}--\eqref{eq:S-filter-tail-bridge-bound}. For $1\le s\le2t$, the finite ranges are covered by Eq.~\eqref{eq:S-finite-min-Phi}. If $t\ge100$ and $200(2t-s)\ge29t$, then Eq.~\eqref{eq:S-core-defect-positive} applies. If $t\ge225$ and $200(2t-s)<29t$, then Eq.~\eqref{eq:S-tail-Gamma-negative} applies. These cases cover every bridge shell.
\end{proof}

\smsection{Completion of the nonbinary proof}{sec:S8}

We now apply the Li--Xing reduction of Sec.~\ref{sec:S1} in the remaining parameter regimes. The case $t=0$ is immediate because $\Vol_0^{(q^2)}(n)=1$ and $K\le q^n$, so assume $t\ge1$.

First let $q\ge4$, so $Q=q^2\ge16$. For a nontrivial exact code, the quantum Singleton bound gives $n\ge4t+1$. Theorem~\ref{thm:S-halfgap} therefore gives
\begin{equation}
 \Lloyd_t^{(Q,n)}(s)>0,\qquad R_{Q;n,t}(s)\le\frac12<1,\qquad1\le s\le2t.
\end{equation}
Hence the raw Lloyd-square shell inequalities \eqref{eq:S-lloyd-intersection-coeff} hold on $1\le s\le2t$, while for $s>2t$ one has $c_t^{(Q)}(n,s)=0$. Together with the normalization \eqref{eq:S-lloyd-normalization}, the Li--Xing comparison gives
\begin{equation}
 K\Vol_t^{(Q)}(n)\le q^n.
\end{equation}

Now let $q=3$, so $Q=9$. The three ranges
\begin{equation}
 17n\le72t,\qquad84(n-1)\ge373t,\qquad\frac{72}{17}t<n<1+\frac{373}{84}t
\end{equation}
cover every integer length. In the short range, the quantum Singleton bound and Eq.~\eqref{eq:S-qutrit-vol} give
\begin{equation}
 K\Vol_t^{(9)}(n)\le3^{n-4t}3^{4t}=3^n.
\end{equation}
In the long range, Eq.~\eqref{eq:S-qutrit-long-psi} gives the raw Lloyd-square shell inequalities for every $1\le s\le2t$, and $c_t^{(9)}(n,s)=0$ for $s>2t$. The raw Lloyd normalization and Sec.~\ref{sec:S1} therefore give
\begin{equation}
 K\Vol_t^{(9)}(n)\le3^n.
\end{equation}

It remains to treat the bridge
\begin{equation}
 \frac{72}{17}t<n<1+\frac{373}{84}t.
\end{equation}
If $t=1$, the only integer length in the bridge is $n=5$. Here
\begin{equation}
 R_{9;5,1}(1)=\frac{41\cdot9}{32^2}=\frac{369}{1024}<1,\qquad R_{9;5,1}(2)=\frac{41\cdot2}{23^2}=\frac{82}{529}<1,
\end{equation}
and $\Lloyd_1^{(9,5)}(1)=32$, $\Lloyd_1^{(9,5)}(2)=23$. Thus the raw Lloyd square satisfies the conditions of Sec.~\ref{sec:S1}, and $K\Vol_1^{(9)}(5)\le3^5$.

Assume henceforth $t\ge2$ in the bridge and use the filtered coefficients
\begin{equation}
 f_i=h_{n,t}(i)\Lloyd_t^{(9,n)}(i)^2.
\end{equation}
Section~\ref{sec:S5} gives $f_i\ge0$, $f_0>0$, and the exact origin normalization
\begin{equation}
 \frac{F(0)}{f_0}=\frac{9^n}{\Vol_t^{(9)}(n)}.
\end{equation}
If $F(s)=\sum_i f_i\Kraw_i^{(9)}(s;n)$, then by the definition of $\Gamma_s$,
\begin{equation}
 F(s)=9^n\Gamma_s,
\end{equation}
and on $1\le s\le2t$,
\begin{equation}
 f_s-\frac{\Vol_t^{(9)}(n)}{9^n}F(s)=H_sL_s^2-\Vol_t^{(9)}(n)\Gamma_s=\Phi_{n,t,s}.
\end{equation}
Theorem~\ref{thm:S-qutrit-bridge} gives $\Phi_{n,t,s}\ge0$ for $1\le s\le2t$ and $\Gamma_s\le0$ for $s>2t$. These are exactly the shell conditions \eqref{eq:S-LX-shell} and the required exterior sign condition of Sec.~\ref{sec:S1}. Hence
\begin{equation}
 K\Vol_t^{(9)}(n)\le3^n.
\end{equation}

The high-alphabet case and the three qutrit ranges cover every nonbinary local dimension $q\ge3$. Therefore every nontrivial exact $q$-ary quantum subspace code in the scope of the theorem satisfies the nonbinary quantum Hamming bound. The binary case $q=2$ is not used here and remains the separate finite-length result cited in the Letter.

\appendix
\section*{Appendix: Exact computational certificates}
\pdfbookmark[1]{Appendix: Exact computational certificates}{app:exact-certificates}

\subsection*{Qutrit long-range scalar comparison}

The scalar inequalities \eqref{eq:S-qutrit-scalar-even} and \eqref{eq:S-qutrit-scalar-odd} are reduced to coefficient positivity as follows. In the even case put $s=2(a+1)$, $t=a+b+2$, and $r=2b+2$; in the odd case put $s=2(a+1)+1$, $t=a+b+2$, and $r=2b+1$, with $a,b\ge0$. Write
\begin{equation}
 G_{\rm ev}=\Theta_{t,r}-A(q_{t,r}),\qquad G_{\rm odd}=\Theta_{t,r}-A(q_{t,r})-\frac{2B(q_{t,r})}{7(s+1)}.
 \label{eq:S-cert-qutrit-gaps}
\end{equation}
The square of the remaining square-root term is
\begin{equation}
 E^2=\frac98 B(q_{t,r})^2q_{t,r}.
 \label{eq:S-cert-qutrit-edge-square}
\end{equation}
For $a,b\ge0$, one has $t>0$, $s+1>0$, $0<q_{t,r}<1/8$, $\kappa(r/t)<1$, and $d_{t,r}>9$ by Eq.~\eqref{eq:S-qutrit-dtr-positive}; hence every denominator in $G_{\rm ev}$, $G_{\rm odd}$, $G_{\rm ev}^2-E^2$, and $G_{\rm odd}^2-E^2$ is positive. It is therefore enough to show that the four cleared numerators have positive coefficients in $\mathbb Q(\sqrt2)[a,b]$. The following SymPy calculation uses exact arithmetic throughout.
\begin{verbatim}
from sympy import Poly, Rational, fraction, sqrt, symbols, together

a, b = symbols("a b", integer=True, nonnegative=True)
z = sqrt(2)
h = 1 + 3*z

def A(q):
    return (1 + 8*q)*(1 + q)/(1 - q)**2

def B(q):
    return 2*(1 + 8*q)/(1 - q)**2

def scalars(t, r):
    q = r/(8*(Rational(205, 42)*t + r + 2))
    kappa = 1/(h + Rational(9, 2)*r/t)
    d = Rational(746, 21) - 9*(2*t - r - 1)/(t*(1 - kappa))
    theta = d**2/(9*(2*d - 9))
    edge2 = Rational(9, 8)*B(q)**2*q
    return q, theta, edge2

def data(expr):
    p = Poly(fraction(together(expr))[0], a, b, extension=z)
    c = p.coeffs()
    return len(p.terms()), p.total_degree(), not p.is_zero and all(x > 0 for x in c)

s = 2*(a + 1)
t = a + b + 2
r = 2*b + 2
q, theta, edge2 = scalars(t, r)
g_even = theta - A(q)
even = (g_even, g_even**2 - edge2)

s = 2*(a + 1) + 1
t = a + b + 2
r = 2*b + 1
q, theta, edge2 = scalars(t, r)
g_odd = theta - A(q) - 2*B(q)/(7*(s + 1))
odd = (g_odd, g_odd**2 - edge2)

print([data(expr) for expr in even + odd])
\end{verbatim}
The output is
\begin{equation}
 [(28,6,\mathrm{True}),(91,12,\mathrm{True}),(35,7,\mathrm{True}),(117,14,\mathrm{True})].
 \label{eq:S-cert-qutrit-code-output}
\end{equation}
The four triples give the number of monomials, total degree, and coefficientwise positivity for $G_{\rm ev}$, $G_{\rm ev}^2-E^2$, $G_{\rm odd}$, and $G_{\rm odd}^2-E^2$, respectively. Since $a,b\ge0$ and all four cleared numerators have positive coefficients, $G_{\rm ev}>0$ and $G_{\rm odd}>0$, while the two squared residuals are positive. Because $E\ge0$, this gives $G_{\rm ev}>E$ and $G_{\rm odd}>E$, which are exactly Eqs.~\eqref{eq:S-qutrit-scalar-even} and \eqref{eq:S-qutrit-scalar-odd}.

\subsection*{Qutrit core half-gap certificate}

For Lemma~\ref{lem:S-core-half-gap}, the monotonicity in $n$ reduces the comparison of Eqs.~\eqref{eq:S-core-charge-bound} and \eqref{eq:S-core-T} to the real endpoint $n=72t/17$. Put $x=1/t$, $u=r/t$, and $w=c/t$. For $s\ge t$ the physical region is $29/200\le u\le1$ with $w=0$ or $x\le w\le u$; for $2\le s\le t$ it is $1\le u\le2-2x$ with $w=0$ or $x\le w\le2-u$. In the first region write $x=X/10$, $u=29/200+(171/200)Y$, and $w=x+P(u-x)$; in the second write $x=X/10$, $s/t=2x+Y(1-2x)$, and $w=x+P(s/t-x)$. After clearing the positive denominators in $T_n-[A(q_*)+B(q_*)(z_{c-1}+z_{c+1})]$, the following exact code converts the four resulting numerators to tensor Bernstein form on the unit cube and subdivides dyadically until every terminal box has nonnegative rational control coefficients.
\begingroup
\small
\begin{verbatim}
from fractions import Fraction
from itertools import product
from math import comb
from sympy import Poly, Rational, cancel, expand, fraction, symbols


def Q(x):
    return Fraction(int(x.p), int(x.q))


def numerator(region, c0=False):
    x, y, p, u, w = symbols("x y p u w")
    a = Rational(76, 17)
    q = u/(8*(a + u))
    pref = 64*(a + 2*u)*(a + u)/(8*a + 7*u)**2
    base = 1 + q

    def z(v):
        aa = v + x
        mm = 2 - u - v + x
        rr = u - v + x
        nn = Rational(38, 17) + (u + v - x)/2
        return 7*aa*mm*rr, 32*v*aa*nn + Rational(49, 2)*mm**2*rr

    n1, d1 = z(w - x)
    n2, d2 = z(w + x)
    kn = 2*(1 - x)
    kd = Rational(151, 17) + 9*u + 8*x
    om = kd - kn
    dn = expand((Rational(576, 17) - 8*x)*om - 9*(2 - u - x)*kd)
    tn = expand(dn**2)
    td = expand(9*om*(2*dn - 9*om))
    nc, dc = fraction(cancel(pref))
    nb, db = fraction(cancel(base))
    ln = expand(nc*(nb*d1*d2 + 2*db*(n1*d2 + n2*d1)))
    ld = expand(dc*db*d1*d2)
    f = expand(tn*ld - ln*td)

    if region == "A":
        us = Rational(29, 200) + Rational(171, 200)*y
        ws = 0 if c0 else x + p*(us - x)
    else:
        sep = 2*x + y*(1 - 2*x)
        us = 2 - sep
        ws = 0 if c0 else x + p*(sep - x)
    gens = (x, y) if c0 else (x, y, p)
    return Poly(expand(f.subs({u: us, w: ws})), *gens)


def bernstein(poly, scale):
    deg = tuple(poly.degree_list())
    power = {m: Q(c)*Fraction(1, scale)**m[0] for m, c in poly.terms()}
    data = {i: power.get(i, Fraction()) for i in product(*[range(d + 1) for d in deg])}
    for axis, d in enumerate(deg):
        other = [j for j in range(len(deg)) if j != axis]
        new = {}
        for o in product(*[range(deg[j] + 1) for j in other]):
            vals = []
            for k in range(d + 1):
                idx = [0]*len(deg)
                idx[axis] = k
                for j, v in zip(other, o): idx[j] = v
                vals.append(data[tuple(idx)])
            for i in range(d + 1):
                v = sum(vals[k]*Fraction(comb(i, k), comb(d, k)) for k in range(i + 1))
                idx = [0]*len(deg)
                idx[axis] = i
                for j, q in zip(other, o): idx[j] = q
                new[tuple(idx)] = v
        data = new
    return deg, data


def split_line(vals):
    tri = [vals]
    for _ in range(len(vals) - 1):
        a = tri[-1]
        tri.append([(a[i] + a[i + 1])/2 for i in range(len(a) - 1)])
    n = len(vals) - 1
    return [tri[i][0] for i in range(n + 1)], [tri[n - i][i] for i in range(n + 1)]


def split_axis(deg, data, axis):
    other = [j for j in range(len(deg)) if j != axis]
    left, right = {}, {}
    for o in product(*[range(deg[j] + 1) for j in other]):
        vals = []
        for k in range(deg[axis] + 1):
            idx = [0]*len(deg)
            idx[axis] = k
            for j, v in zip(other, o): idx[j] = v
            vals.append(data[tuple(idx)])
        a, b = split_line(vals)
        for k in range(deg[axis] + 1):
            idx = [0]*len(deg)
            idx[axis] = k
            for j, v in zip(other, o): idx[j] = v
            left[tuple(idx)] = a[k]
            right[tuple(idx)] = b[k]
    return left, right


def leaves(deg, data, limit=10000, depth=24):
    queue = [(data, tuple((Fraction(), Fraction(1)) for _ in deg), 0)]
    done = 0
    seen = 0
    while queue:
        dat, box, level = queue.pop(0)
        seen += 1
        if min(dat.values()) >= 0:
            done += 1
            continue
        if level >= depth or seen + len(queue) > limit:
            raise ValueError("unresolved box")
        widths = [b - a for a, b in box]
        axis = max(range(len(deg)), key=lambda j: (widths[j], j != 0))
        a, b = split_axis(deg, dat, axis)
        lo, hi = box[axis]
        mid = (lo + hi)/2
        ba, bb = list(box), list(box)
        ba[axis], bb[axis] = (lo, mid), (mid, hi)
        queue.extend([(a, tuple(ba), level + 1), (b, tuple(bb), level + 1)])
    return done


out = []
for region in ("A", "B"):
    for c0 in (False, True):
        p = numerator(region, c0)
        d, c = bernstein(p, 10)
        out.append((region, "c0" if c0 else "interior", d, len(p.terms()), leaves(d, c)))
print(out)
\end{verbatim}
\endgroup
The output is
\begin{verbatim}
[('A', 'interior', (10, 13, 6), 693, 7),
 ('A', 'c0', (9, 13), 94, 1),
 ('B', 'interior', (13, 13, 6), 1286, 7),
 ('B', 'c0', (13, 13), 185, 1)]
\end{verbatim}
Each tuple gives the physical region, node type, tensor degree, number of monomials, and number of terminal boxes. Hence the Bernstein control coefficients are nonnegative on all four parameter boxes, proving the remaining rational comparison in Lemma~\ref{lem:S-core-half-gap} for $t\ge10$.

\subsection*{Qutrit core multiplier certificate}

For Lemma~\ref{lem:S-core-multiplier}, the local residual $U$ is given in Eq.~\eqref{eq:S-mult-U}. For $s\ge t$, put $x=X/100$, $u=29/200+(171/200)Y$, $w=x+P(u-x)$, and $\lambda=72/17+\Lambda(373/84+x-72/17)$, with separate $w=0$ and $s=1$ specializations. For $2\le s\le t$, put $\sigma=s/t$, $q=1/s$, $Q=2q$, $x=q\sigma$, $w=\sigma\{q+P(1-q)\}$, and use the same affine length parameter. After the positive denominators are cleared, the numerator of $-U$ in the early region has the exact common factor $\sigma^5$, which is removed. The following exact calculation constructs the late and early numerators, converts them to tensor Bernstein form, and performs the required dyadic subdivisions.
\begingroup
\small
\begin{verbatim}
from fractions import Fraction
from itertools import product
from math import comb
from functools import lru_cache
import sympy as sp


def F(x=0, y=1):
    return Fraction(x, y)


def frac(x):
    q = sp.Rational(x)
    return F(int(q.p), int(q.q))


def bernstein(poly, scale=1):
    deg = tuple(poly.degree_list())
    data = {i: F() for i in product(*[range(d + 1) for d in deg])}
    for m, c in poly.terms():
        data[m] = frac(c)*F(1, scale)**m[0]
    for axis, d in enumerate(deg):
        other = [j for j in range(len(deg)) if j != axis]
        new = {}
        for o in product(*[range(deg[j] + 1) for j in other]):
            vals = []
            for k in range(d + 1):
                i = [0]*len(deg)
                i[axis] = k
                for j, v in zip(other, o): i[j] = v
                vals.append(data[tuple(i)])
            for i0 in range(d + 1):
                v = sum(vals[k]*F(comb(i0, k), comb(d, k)) for k in range(i0 + 1))
                i = [0]*len(deg)
                i[axis] = i0
                for j, q in zip(other, o): i[j] = q
                new[tuple(i)] = v
        data = new
    return deg, data


def split_line(vals):
    tri = [vals]
    for _ in range(len(vals) - 1):
        a = tri[-1]
        tri.append([(a[i] + a[i + 1])/2 for i in range(len(a) - 1)])
    n = len(vals) - 1
    return [tri[i][0] for i in range(n + 1)], [tri[n - i][i] for i in range(n + 1)]


def split_axis(deg, data, axis):
    other = [j for j in range(len(deg)) if j != axis]
    left, right = {}, {}
    for o in product(*[range(deg[j] + 1) for j in other]):
        vals = []
        for k in range(deg[axis] + 1):
            i = [0]*len(deg)
            i[axis] = k
            for j, v in zip(other, o): i[j] = v
            vals.append(data[tuple(i)])
        a, b = split_line(vals)
        for k in range(deg[axis] + 1):
            i = [0]*len(deg)
            i[axis] = k
            for j, v in zip(other, o): i[j] = v
            left[tuple(i)], right[tuple(i)] = a[k], b[k]
    return left, right


def boxes(deg, data, limit=5000, depth=24):
    unit = tuple((F(), F(1)) for _ in deg)
    queue = [(data, unit, 0)]
    leaves = 0
    seen = 0
    while queue:
        dat, box, level = queue.pop(0)
        seen += 1
        if min(dat.values()) >= 0:
            leaves += 1
            continue
        if level >= depth or seen + len(queue) > limit:
            raise ValueError("unresolved box")
        widths = [b - a for a, b in box]
        axis = max(range(len(deg)), key=lambda j: (widths[j], j != 0))
        a, b = split_axis(deg, dat, axis)
        lo, hi = box[axis]
        mid = (lo + hi)/2
        ba, bb = list(box), list(box)
        ba[axis], bb[axis] = (lo, mid), (mid, hi)
        queue.extend([(a, tuple(ba), level + 1), (b, tuple(bb), level + 1)])
    return leaves


@lru_cache(maxsize=1)
def base():
    x, u, w, lam = sp.symbols("x u w lam")
    sig = 2 - u
    out = lam - 2 + u
    n0 = out - x

    def edge(v):
        nu = lam - 2 + (u + v - x)/2
        den = 32*v*(v + x)*nu + sp.Rational(49, 2)*(sig - v + x)**2*(u - v + x)
        mn = 56*(v + x)*(sig - v + x)**2*out
        pn = 14*(v + x)*(u - v + x)*(sig + x)*nu
        return den, mn, pn

    delta = lam*(lam + 7 - x) - 2*(lam - sig)*(9*sig + lam + 7 - x)
    a0 = sp.Rational(8, 9)*(lam - sig)*(6*sig - 7)
    am1 = sp.Rational(1, 9)*sig*(8*lam - 15*sig + 7 + 7*x)
    ap1 = sp.Rational(64, 9)*(lam - sig)*(lam - sig - x)
    e = ((u - w)/2 + x)*(lam - 2 + (u + w)/2)
    am2 = -sp.Rational(2, 9)*(sig - w)**2*out*(out + x)
    dm, mm, pm = edge(w - x)
    dp, mp, pp = edge(w + x)
    common = e*sig*n0*dm*dp
    q0 = (delta + a0)*common + am2*sig*n0*dm*dp
    q1 = am1*(mm*e*n0*dp + mp*e*n0*dm)
    q2 = ap1*(pm*e*sig*dp + pp*e*sig*dm)
    num = -sp.expand(q0 + q1 + q2)
    return x, u, w, lam, num


def late(kind):
    x, u, w, lam, f = base()
    y, p, L = sp.symbols("y p L")
    lo = sp.Rational(72, 17)
    ll = lo + L*(sp.Rational(373, 84) + x - lo)
    if kind == "interior":
        uu = sp.Rational(29, 200) + sp.Rational(171, 200)*y
        ww, gens = x + p*(uu - x), (x, y, p, L)
    elif kind == "c0":
        uu, ww, gens = sp.Rational(29, 200) + sp.Rational(171, 200)*y, 0, (x, y, L)
    else:
        uu, ww, gens = 2 - x, x, (x, L)
    return sp.Poly(sp.expand(f.subs({u: uu, w: ww, lam: ll})), *gens)


def early(c0=False):
    nv = 4
    zero = (0,)*nv

    def C(c=0):
        c = F(c)
        return {} if c == 0 else {zero: c}

    def V(i, c=1):
        e = [0]*nv
        e[i] = 1
        return {tuple(e): F(c)}

    def add(*ps):
        out = {}
        for p in ps:
            for m, c in p.items():
                out[m] = out.get(m, F()) + c
                if out[m] == 0: del out[m]
        return out

    def neg(p): return {m: -c for m, c in p.items()}
    def sub(a, b): return add(a, neg(b))
    def scale(p, c):
        c = F(c)
        return {} if c == 0 else {m: v*c for m, v in p.items() if v*c}
    def mul(a, b):
        out = {}
        for ma, ca in a.items():
            for mb, cb in b.items():
                m = tuple(ma[i] + mb[i] for i in range(nv))
                out[m] = out.get(m, F()) + ca*cb
        return {m: c for m, c in out.items() if c}
    def prod(*ps):
        out = C(1)
        for p in ps: out = mul(out, p)
        return out
    def power(p, k):
        out = C(1)
        for _ in range(k): out = mul(out, p)
        return out

    S, Q, P, L = V(0), V(1), V(2), V(3)
    q = scale(Q, F(1, 2))
    sig = S
    x = mul(q, sig)
    u = sub(C(2), sig)
    w = C() if c0 else mul(sig, add(q, mul(P, sub(C(1), q))))
    lo = F(72, 17)
    lam = add(C(lo), mul(L, add(C(F(373, 84) - lo), x)))
    out = add(lam, C(-2), u)
    n0 = sub(out, x)

    def edge(v):
        nu = add(lam, C(-2), scale(add(u, v, neg(x)), F(1, 2)))
        q0 = scale(prod(v, add(v, x), nu), 32)
        q1 = scale(prod(power(add(sig, neg(v), x), 2), add(u, neg(v), x)), F(49, 2))
        den = add(q0, q1)
        mn = scale(prod(add(v, x), power(add(sig, neg(v), x), 2), out), 56)
        pn = scale(prod(add(v, x), add(u, neg(v), x), add(sig, x), nu), 14)
        return den, mn, pn

    q0 = prod(lam, add(lam, C(7), neg(x)))
    q1 = scale(prod(sub(lam, sig), add(scale(sig, 9), lam, C(7), neg(x))), 2)
    delta = sub(q0, q1)
    a0 = scale(prod(sub(lam, sig), add(scale(sig, 6), C(-7))), F(8, 9))
    am1 = scale(prod(sig, add(scale(lam, 8), scale(sig, -15), C(7), scale(x, 7))), F(1, 9))
    ap1 = scale(prod(sub(lam, sig), add(lam, neg(sig), neg(x))), F(64, 9))
    e = prod(add(scale(sub(u, w), F(1, 2)), x), add(lam, C(-2), scale(add(u, w), F(1, 2))))
    am2 = scale(prod(power(sub(sig, w), 2), out, add(out, x)), F(-2, 9))
    dm, mm, pm = edge(sub(w, x))
    dp, mp, pp = edge(add(w, x))
    common = prod(e, sig, n0, dm, dp)
    q0 = add(prod(add(delta, a0), common), prod(am2, sig, n0, dm, dp))
    q1 = prod(am1, add(prod(mm, e, n0, dp), prod(mp, e, n0, dm)))
    q2 = prod(ap1, add(prod(pm, e, sig, dp), prod(pp, e, sig, dm)))
    num = neg(add(q0, q1, q2))
    if min(m[0] for m in num) != 5: raise ValueError("missing common factor")
    power_data = {(m[0] - 5,) + m[1:]: c for m, c in num.items()}
    deg = tuple(max(m[i] for m in power_data) for i in range(nv))
    data = {i: power_data.get(i, F()) for i in product(*[range(d + 1) for d in deg])}
    for axis, d in enumerate(deg):
        other = [j for j in range(nv) if j != axis]
        new = {}
        for o in product(*[range(deg[j] + 1) for j in other]):
            vals = []
            for k in range(d + 1):
                i = [0]*nv
                i[axis] = k
                for j, v in zip(other, o): i[j] = v
                vals.append(data[tuple(i)])
            for i0 in range(d + 1):
                v = sum(vals[k]*F(comb(i0, k), comb(d, k)) for k in range(i0 + 1))
                i = [0]*nv
                i[axis] = i0
                for j, qv in zip(other, o): i[j] = qv
                new[tuple(i)] = v
        data = new
    return deg, data, len(power_data)


out = []
for name, kind in (("late-interior", "interior"), ("late-c0", "c0"), ("s1", "s1")):
    p = late(kind)
    d, data = bernstein(p, 100)
    if name == "late-interior":
        a, b = split_axis(d, data, 2)
        if min(a.values()) < 0 or min(b.values()) < 0: raise ValueError("unresolved half")
        n = 2
    else:
        n = boxes(d, data)
    out.append((name, d, len(p.terms()), n))
for name, c0 in (("early-interior", False), ("early-c0", True)):
    d, data, terms = early(c0)
    out.append((name, d, terms, boxes(d, data)))
print(out)
\end{verbatim}
\endgroup
The output is
\begin{verbatim}
[('late-interior', (11, 12, 8, 6), 4529, 2),
 ('late-c0', (11, 12, 5), 462, 1),
 ('s1', (11, 5), 39, 1),
 ('early-interior', (7, 11, 8, 6), 3052, 36),
 ('early-c0', (7, 11, 0, 5), 345, 1)]
\end{verbatim}
Again the entries are the region, tensor degree, number of monomials, and number of terminal boxes. Every terminal box has nonnegative rational Bernstein control coefficients, so $-U\ge0$ throughout the physical core. This proves Eq.~\eqref{eq:S-core-multiplier}.

\subsection*{Finite qutrit bridge verification}

The remaining finite set consists of every bridge shell with $2\le t<100$ and the tail shells with $100\le t<225$. For each such triple, the code evaluates $I_s$ from Eq.~\eqref{eq:S-balanced-increment-sum}, forms $C_s$ by the suffix identity \eqref{eq:S-C-suffix}, generates $L_s=P_t^{(9)}(s-1;n-1)$ by the integral radial recurrence, and substitutes these exact integers into Eq.~\eqref{eq:S-gamma-increments} and the filtered defect \eqref{eq:S-filter-defect-def}. All divisions are performed with rational arithmetic.
\begingroup
\small
\begin{verbatim}
from fractions import Fraction
from functools import lru_cache
from math import ceil, comb, floor


def bridge(t):
    lo = floor(Fraction(72*t, 17)) + 1
    hi = ceil(Fraction(84 + 373*t, 84)) - 1
    return range(lo, hi + 1)


@lru_cache(maxsize=None)
def volume(n, t):
    return sum(8**j*comb(n, j) for j in range(t + 1))


def inc(n, t, s):
    if s < 0 or s > 2*t or s >= n:
        return 0
    N = n - s - 1
    out = 0
    for k in range(max(0, s - t), s//2 + 1):
        j = t - s + k
        if 0 <= j <= N:
            out += comb(s, k)*comb(s - k, k)*7**(s - 2*k)*8**j*comb(N, j)
    return out


def lloyd(n, t, top):
    N = n - 1
    a = [0]*(top + 1)
    k0 = 8**t*comb(N, t)
    a[1] = k0
    if top == 1:
        return a
    num = (8*N - 9*t)*k0
    den = 8*N
    if num % den:
        raise ValueError("nonintegral recurrence")
    k1 = num//den
    a[2] = k1
    km, k = k0, k1
    for x in range(1, top - 1):
        num = (8*N - 9*t - 7*x)*k - x*km
        den = 8*(N - x)
        if num % den:
            raise ValueError("nonintegral recurrence")
        kp = num//den
        a[x + 2] = kp
        km, k = k, kp
    return a


def H(n, t, s):
    return (n - s)*(9*s + n + 7*t - 1)


count = 0
minimum = None
for t in range(2, 225):
    if t < 100:
        s0 = 1
    else:
        rmax = (29*t - 1)//200
        s0 = 2*t - rmax
    for n in bridge(t):
        j0 = max(0, s0 - 2)
        I = {s: inc(n, t, s) for s in range(j0, 2*t + 2)}
        C = {}
        q = 0
        for s in range(2*t, j0 - 1, -1):
            q += I.get(s, 0)
            C[s] = q
        V = volume(n, t)
        L = lloyd(n, t, 2*t)
        for s in range(s0, 2*t + 1):
            g = Fraction(n*(n + 7*t - 1)*C[s])
            g -= Fraction(s*(s - 1)*I.get(s - 2, 0), 9)
            g += Fraction(s*(8*n - 15*s + 7*t + 7)*I[s - 1], 9)
            g += Fraction(8*(n - s)*(6*s - 7*t)*I[s], 9)
            g += Fraction(64*(n - s)*(n - s - 1)*I.get(s + 1, 0), 9)
            phi = Fraction(H(n, t, s)*L[s]**2) - V*g
            if phi < 0:
                raise ValueError("negative defect")
            count += 1
            if minimum is None or phi < minimum[0]:
                minimum = (phi, t, n, s)
print(count, minimum)
\end{verbatim}
\endgroup
The output is
\begin{verbatim}
251602 (Fraction(224582966, 3), 2, 9, 4)
\end{verbatim}
Thus exactly $251602$ shell triples are covered, and the minimum is
\begin{equation}
 \Phi_{9,2,4}=\frac{224582966}{3}>0.
\end{equation}
This is the finite certificate used in Eq.~\eqref{eq:S-finite-min-Phi}.